\documentclass[%
 reprint,
superscriptaddress,
 amsmath,amssymb,
 aps,
twocolumn,
]{revtex4-2}
\usepackage{xcolor}
\usepackage{graphicx}% Include figure files
\usepackage{dcolumn}% Align table columns on decimal point
\usepackage{bm}% bold math
\usepackage[colorlinks,allcolors=blue,urlcolor=blue]{hyperref}
\usepackage[normalem]{ulem}

\begin{document}

\preprint{APS/123-QED}

\title{Advancing Simulations of Coupled Electron and Phonon Nonequilibrium Dynamics\\ Using Adaptive and Multirate Time Integration}

\author{Jia Yao}
\affiliation{
 Department of Applied Physics and Materials Science, and Department of Physics, California Institute of Technology, Pasadena, California 91125
}

\author{Ivan Maliyov}
\affiliation{Mathematics for Materials Modeling, Institute of Mathematics and Institute of Materials, EPFL, CH-1015 Lausanne, Switzerland}

\author{David J. Gardner} % ORCID: 0000-0002-7993-8282
\affiliation{
 Center for Applied Scientific Computing, Lawrence Livermore National Laboratory, Livermore, California 94550
}

\author{Carol S. Woodward} % ORCID: 0000-0002-6502-8659
\affiliation{
 Center for Applied Scientific Computing, Lawrence Livermore National Laboratory, Livermore, California 94550
}

\author{Marco Bernardi\thanks{}}\email{bmarco@caltech.edu}%
\affiliation{
 Department of Applied Physics and Materials Science, and Department of Physics, California Institute of Technology, Pasadena, California 91125
}

\begin{abstract}
% summary of work
%\noindent
Electronic structure calculations in the time domain provide a deeper understanding of nonequilibrium dynamics in materials. The real-time Boltzmann equation (rt-BTE), used in conjunction with accurate interactions computed from first principles, has enabled reliable predictions of coupled electron and lattice dynamics. 
However, the timescales and system sizes accessible with this approach are still limited, with two main challenges being the different timescales of electron and phonon interactions and the cost of computing collision integrals. 
As a result, only a few examples of these calculations exist, mainly for two-dimensional (2D) materials. 
Here we leverage adaptive and multirate time integration methods to achieve a major step forward in solving the coupled rt-BTEs for electrons and phonons. Relative to conventional (non-adaptive) time-stepping, our approach achieves a 10x speedup for a given target accuracy, or greater accuracy by 3--6 orders of magnitude for the same computational cost, enabling efficient calculations in both 2D and bulk materials.
This efficiency is showcased by computing the coupled electron and lattice dynamics in graphene up to $\sim$100 ps, as well as modeling ultrafast lattice dynamics and thermal diffuse scattering maps in a bulk material (silicon). 
These results open new opportunities for quantitative studies of nonequilibrium physics in materials. 
\end{abstract}
\maketitle

\section{Introduction}
\label{s:intro}
%\vspace{-10pt}
Coherent control of electronic and lattice degrees of freedom is a novel frontier in materials physics~\cite{disa2021engineering,basov2017towards}.
\mbox{Experiments} have shown coherent phonon excitation~\cite{Murray-bismuth2005, Forst-control2011, ruello-review2015},  
tuned physical properties and interactions at ultrafast timescales~\cite{Fausti2011light, Mitrano-Sc2016,Afanasiev-control2021}, and achieved light-induced phase transitions~\cite{Cavalleri-phase2004, Kim-sdw2012, Hellmann-cdw2012}.  
These dynamics can be probed experimentally using techniques such as transient absorption and reflectivity~\cite{Katz-2012xc, Zurch-germanium2017, Cushing-hc2018}, inelastic X-ray scattering~\cite{Mohr-inelaticxray2007, Trigo-xray2013}, time- and angle-resolved photoemission spectroscopy~\cite{Valla-arpes1999, Na-graphene2019} and transient thermal gratings~\cite{Maznev-thermal2011}, among others. 
In parallel, advances in computing electron and phonon interactions from first principles~\cite{Baroni-DFT2001, bernardi-theory2016,Zhou-perturbo2021,Giustino-Review2017} have enabled quantitative simulations of nonequilibrium phenomena such as excited carrier relaxation~\cite{Bernardi-thermalization2014, Sangalli-photo2015, Jhalani-BTE2017, sjakste2018hot, Zheng-theory2023}, high-field transport~\cite{Maliyov-BTE2021}, ultrafast spectroscopies~\cite{maliyov2024dynamic} and more recently simulations of coupled electron and phonon dynamics~\cite{Tong-BTE2021, Caruso-BTE2021, Caruso-chiral}.
These methods can unravel microscopic mechanisms of ultrafast dynamics and quantitatively interpret time-domain spectroscopies.

Various first-principles methodologies address distinct regimes of nonequilibrium dynamics, each tailored to specific time and interaction scales.
Nonadiabatic molecular dynamics~\cite{Prezhdo-NAMD2021,Zheng-theory2023} and nonequilibrium Green's functions~\cite{Perfetto-GW2022, Perfetto2023real} are typically employed to model femtosecond electron dynamics.
Time-dependent density functional theory can also effectively model coherent femtosecond dynamics of electrons following optical excitations~\cite{Rozzi}, but capturing picosecond phonon dynamics remains difficult.

This work focuses on the real-time Boltzmann transport equation (rt-BTE) method~\cite{Jhalani-BTE2017, Zhou-perturbo2021, Maliyov-BTE2021, Tong-BTE2021} with first-principles electron and phonon interactions. 
The rt-BTE is a set of coupled integro-differential equations for the time-dependent electron and phonon occupations in momentum space, hereon referred to as populations.
Both the electron-phonon ($e$-ph) and phonon-phonon (ph-ph) scattering integrals are computed on dense momentum grids at each time step, with timescales of interest ranging from femtoseconds (fs) for electrons to hundreds of picoseconds (ps) for phonons. Using dense momentum grids ensures precise resolution of scattering processes and energy conservation, which are critical for reliable predictions of the dynamics.
The ability to capture anharmonic ph-ph interactions~\cite{Maradudin-anharm1962, Cowley-anharm1968, Debernardi-anharm1995, Li-ShengBTE2014, Atsushi-phono3py2015, carrete-almaBTE2017} is particularly important for modeling nonequilibrium lattice dynamics at the ps timescale and in materials with strong \mbox{anharmonicity.}

However, due to computational challenges with propagating the rt-BTE in time, only a handful of recent works have shown first-principles simulations of phonon dynamics in the time domain~\cite{Caruso-BTE2021, Tong-BTE2021, Britt-BTE2022}.
Two key challenges are the different timescales of electron and phonon dynamics and the computational cost of evaluating collision integrals, particularly for ph-ph interactions. 
Typically, the rt-BTE is evolved with a fixed time step of a few fs to resolve the faster electron dynamics, while the simulation extends to tens of ps for the phonons to reach steady state or thermal equilibrium.
At each time step, the ph-ph scattering integral is computed by integrating over a dense momentum grid in the Brillouin zone (BZ) for all phonon modes. 
In bulk systems, this requires summing over billions of scattering channels, which is several orders of magnitude more expensive than evaluating the $e$-ph scattering integrals. 
Therefore, a 10-ps simulation for a 2D material with a few atoms in the unit cell can easily exceed thousands of CPU core hours on modern high-performance computers~\cite{Tong-BTE2021}.
Without a significant reduction in computational cost, simulations of coupled electron and phonon dynamics remain out of reach for bulk materials and for all but the simplest 2D materials.

To address these challenges, we explore time integration methods from the \textsc{sundials} suite \cite{hindmarsh2005sundials, gardner2022enabling}, which offers an array of robust and efficient algorithms for integrating differential equations.
The \textsc{arkode} package \cite{reynolds2023arkode} within \textsc{sundials} contains adaptive step size Runge-Kutta (RK) \cite{hairer2008solving} as well as multirate infinitesimal (MRI) methods \cite{sandu2019class}. 
Adaptive RK methods dynamically adjust the time step to obtain solutions within user-specified solution tolerances while maximizing efficiency, selecting step sizes that reflect the inherent dynamics of the system.
MRI methods further improve efficiency by time-evolving different processes or solution components with different step sizes. This is particularly effective when the dynamics is governed by processes with well-separated timescales and the slower processes dominates the computational cost, making it an ideal fit for the rt-BTE. 
Methods in the MRI family have demonstrated improved efficiency in air pollution models \cite{schlegel2009multirate} and reactive flow simulations \cite{loffeld2024performance} compared to single-rate integrators and multirate spectral deferred correction methods, respectively. 
Despite their success in these areas, their application to electron and lattice dynamics simulated from first principles remains \mbox{unexplored.}

In this work, we use adaptive RK and MRI time integration methods to simulate coupled electron and phonon dynamics in the rt-BTE framework, implementing an interface between the \textsc{Perturbo} code~\cite{Zhou-perturbo2021} and the \textsc{sundials} library~\cite{hindmarsh2005sundials, gardner2022enabling}. 
We achieve a significant reduction in computational cost while retaining the same accuracy by using different and adaptive time steps for $e$-ph and ph-ph interactions. %compared to the 4th-order Runge Kutta (RK4) method. 
We benchmark our approach on graphene, achieving a reduction in computational cost by orders of magnitude for any choice of solution tolerance. 
%% DO WE SHOW THIS ANALYSIS?
Finally, our adaptive and multirate time-stepping scheme allows us to solve the coupled rt-BTE for bulk materials with reasonable computational effort, a goal that has so far been considered out of reach. %milestone
This is demonstrated by studying nonequilibrium lattice dynamics in silicon and simulating the associated thermal diffuse scattering maps.
Taken together, the methods developed in this work advance accurate simulations of nonequilibrium physics in real materials, opening doors for future studies of ultrafast electronic and lattice dynamics in materials driven by optical or terahertz pulses. %and ultrafast devices.

\section{Results}
\label{s:results}
\subsection{Real-time Boltzmann transport equation}
\begin{figure*}[t]
\centering
\includegraphics[width=1.0\textwidth]{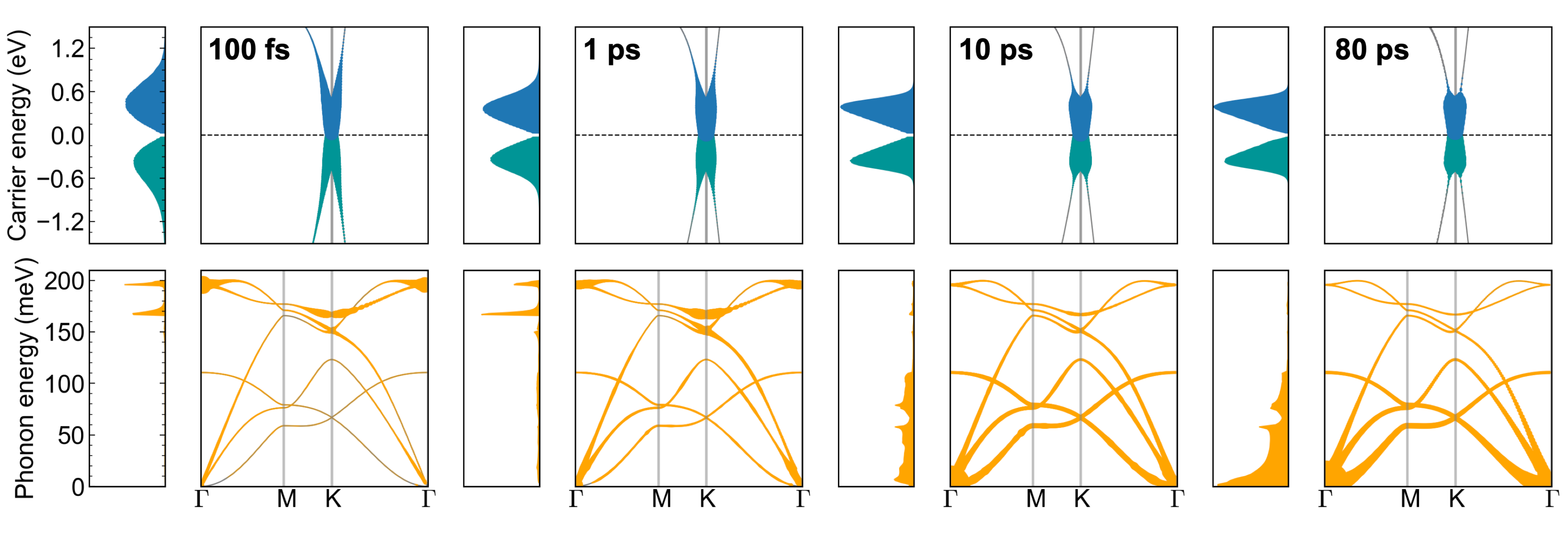}
\caption{\textbf{Coupled electron and phonon dynamics in graphene}. 
Top row: electron populations, $f_{n\textbf{k}}(t)$ (blue), and hole populations, $1-f_{n\textbf{k}}(t)$ (green), mapped onto the band structure at different time snapshots labeled in the panels, with point sizes proportional to the populations. To the left of each panel, we show the carrier populations as a function of energy by averaging over the BZ. 
Bottom row: change in phonon population, $\Delta N_{\nu\textbf{q}}(t)$, relative to the initial distribution, shown on the phonon dispersion. The BZ-average is also shown to the left of each phonon dispersion plot. At time $t=0$, the initial carriers are set to a hot Fermi-Dirac distribution at $4000$~K, while the phonons follow a Bose-Einstein distribution at $300$~K.}
\label{fig1}
\end{figure*}

Using the \textsc{perturbo}~\cite{Tong-BTE2021, Zhou-perturbo2021} code, we propagate in time the coupled electron and phonon rt-BTEs in a homogeneously excited material:
\begin{equation}\label{eq:bte}
\begin{split}
    \frac{\partial f_{n\mathbf{k}}(t)}{\partial t} &= \mathcal{I}^{e\mathrm{\text{-}ph}}[f_{n\mathbf{k}}(t),N_{\nu\mathbf{q}}(t)], \\
    \frac{ \partial N_{\nu\mathbf{q}}(t)}{\partial t}&= \mathcal{I}^{\mathrm{ph}\text{-}e}[f_{n\mathbf{k}}(t),N_{\nu\mathbf{q}}(t)] + \mathcal{I}^{\mathrm{ph\text{-}ph}}[N_{\nu\mathbf{q}}(t)].
\end{split}
\end{equation}
The electron populations, $f_{n\mathbf{k}}(t)$, are labeled by the electronic band index $n$ and crystal momentum $\mathbf{k}$ (hole carriers can be simulated with populations $1-f_{n\mathbf{k}}(t)$), while the phonon populations, $N_{\nu\mathbf{q}}(t)$, are labeled by the mode index $\nu$ and wave-vector $\mathbf{q}$. 
In Eq.~\eqref{eq:bte}, the $e$-ph  scattering integrals, $\mathcal{I}^{e\text{-}\mathrm{ph}}$ and $\mathcal{I}^{\mathrm{ph}\text{-}e}$, and ph-ph scattering integral, $\mathcal{I}^{\mathrm{ph\text{-}ph}}$, are computed from the corresponding interaction matrix elements (see Methods). 

Previous work has simulated coupled carrier and phonon dynamics in graphene by advancing the rt-BTE using the 4th-order Runge-Kutta (RK4) method with a fixed time step size~\cite{Tong-BTE2021}.
While the $e$-ph and ph-ph scattering matrices can be computed and stored before the real-time simulation, the scattering integrals need to be recomputed at each time step and require integration over the BZ (see Methods). 
Computing the ph-ph scattering integrals takes about two orders of magnitude more runtime than the $e$-ph integrals due to differences in the scattering processes in momentum space.
At each time step, the computational cost of evaluating the $e$-ph scattering integrals scales as $\mathcal{N}_\mathrm{c}\,\mathcal{N}_\mathrm{ph}$, where $\mathcal{N}_\mathrm{c}$ is the number of carrier momenta and band indices (labels $n\textbf{k}$), and $\mathcal{N}_\mathrm{ph}$ is the number of phonon momenta and mode indices (labels $\nu\textbf{q}$).
In contrast, the computational cost for the ph-ph scattering integrals scales as $\mathcal{N}^2_\mathrm{ph}$.
In the low-excitation regime, the carriers can be sampled in a few eV energy windows near the band edge or Fermi energy, and thus in practice $\mathcal{N}_\mathrm{ph} \gg \mathcal{N}_\mathrm{c}$.
This shows that computing the ph-ph collision integral $\mathcal{I}^{\mathrm{ph\text{-}ph}}$ and the phonon dynamics is significantly more expensive -- in practical calculations by 1--2 orders of magnitude -- than computing the dynamics governed by $e$-ph interactions.

\subsection{Adaptive and multirate time integration}
%\vspace{-10pt} 
To improve the computational efficiency, we explore adaptive time-stepping methods, which can adjust the time step size according to the underlying dynamics. Motivated by the different timescales for carrier (fs) and phonon (ps) dynamics, we also apply multirate methods that use different time steps for $e$-ph and ph-ph scattering integrals. 

Adaptive-step explicit Runge-Kutta (ERK) methods target ordinary differential equations (ODEs), %of the form
\begin{align} \label{eq:erk_ode}
    y' &= f(t, y),  \quad y(t_0) = y_0.
\end{align}
For the coupled rt-BTEs, the solution vector $y$ contains all carrier and phonon populations in the respective momentum grids:
\begin{equation} \label{eq:state_vec}
    y(t) =
    \begin{bmatrix}
    f_{n\textbf{k}}(t) \\
    N_{\nu\textbf{q}}(t)
    \end{bmatrix}, 	
\end{equation}
and the right-hand side (RHS) function in Eq.~\ref{eq:erk_ode} is
\begin{equation} \label{eq:erk_rhs}
    f(t, y) =
    \begin{bmatrix}
    \mathcal{I}^{e\text{-}\mathrm{ph}}(y) \\
    \mathcal{I}^{\mathrm{ph}\text{-}e}(y) + \mathcal{I}^{\mathrm{ph\text{-}ph}}(y)
    \end{bmatrix}.
\end{equation}
With an adaptive integration method, the time step is adjusted to satisfy user-defined error tolerances, the choice of which is critical to adaptive integrator performance. 
To solve the rt-BTE with adaptive time-stepping, we specify the relative and absolute tolerances for carriers and phonons in \textsc{sundials} (see  Methods).

MRI methods target ODEs with the RHS function split into fast and slow parts, 
\begin{align} \label{eq:mri_ode}
    y' &= f^{f}(t, y)+ f^{s}(t, y), \quad y(t_0) = y_0,
\end{align}
where $e$-ph interactions are considered the fast part and ph-ph interactions the slow component:
\begin{equation} \label{eq:mri_rhs}
    f^{f}(t, y) =
    \begin{bmatrix}
    \mathcal{I}^{e\text{-}\mathrm{ph}}(y) \\
    \mathcal{I}^{\mathrm{ph}\text{-}e}(y)
    \end{bmatrix}, \quad	
    f^{s}(t, y) =
    \begin{bmatrix}
    0 \\
    \mathcal{I}^{\mathrm{ph\text{-}ph}}(y)
    \end{bmatrix}.
\end{equation}
MRI methods advance the slow dynamics at a fixed time step, $h_s$, while the fast dynamics are evolved by solving an auxiliary ODE with an adaptive time step using a sufficiently accurate method such as an adaptive ERK method. Both time integration approaches are available in the \textsc{arkode} package in \textsc{sundials} and are accessed through our interface in \textsc{Perturbo}.

\subsection{Electron and phonon dynamics in graphene}

%%% description of figure 1
Leveraging adaptive and multirate time integration methods, we study the coupled dynamics of carriers and phonons in graphene, where evolving the system for long times, up to tens of ps to access lattice dynamics, is computationally expensive.
The initial state of excited carriers is set to a hot Fermi-Dirac distribution at 4000 K~\cite{Tong-BTE2021} while the phonons are set to a Bose-Einstein distribution at 300 K.

Using a third-order MRI method, we are able to extend the simulations to 80 ps with a reasonable computational cost.
Fig.~\ref{fig1} shows the carrier and phonon populations along a high-symmetry path at various time snapshots, capturing the main trends of the coupled dynamics.
In the first ps, electrons and holes undergo rapid cooling and emit $\mathrm{A'_1}$ and $\mathrm{E_{2g}}$ optical phonons with momenta near $\Gamma$ and $\mathrm{K}$, due to the strong $e$-ph couplings with these phonons~\cite{Na-graphene2019}.
From 1 to 10 ps, the excess optical phonons decay into acoustic phonons near $\Gamma$.
The thermalization process in the ZA branch is slow and extends to more than 80 ps.

\begin{figure}[t]
\centering
\includegraphics[width=1.0\columnwidth]{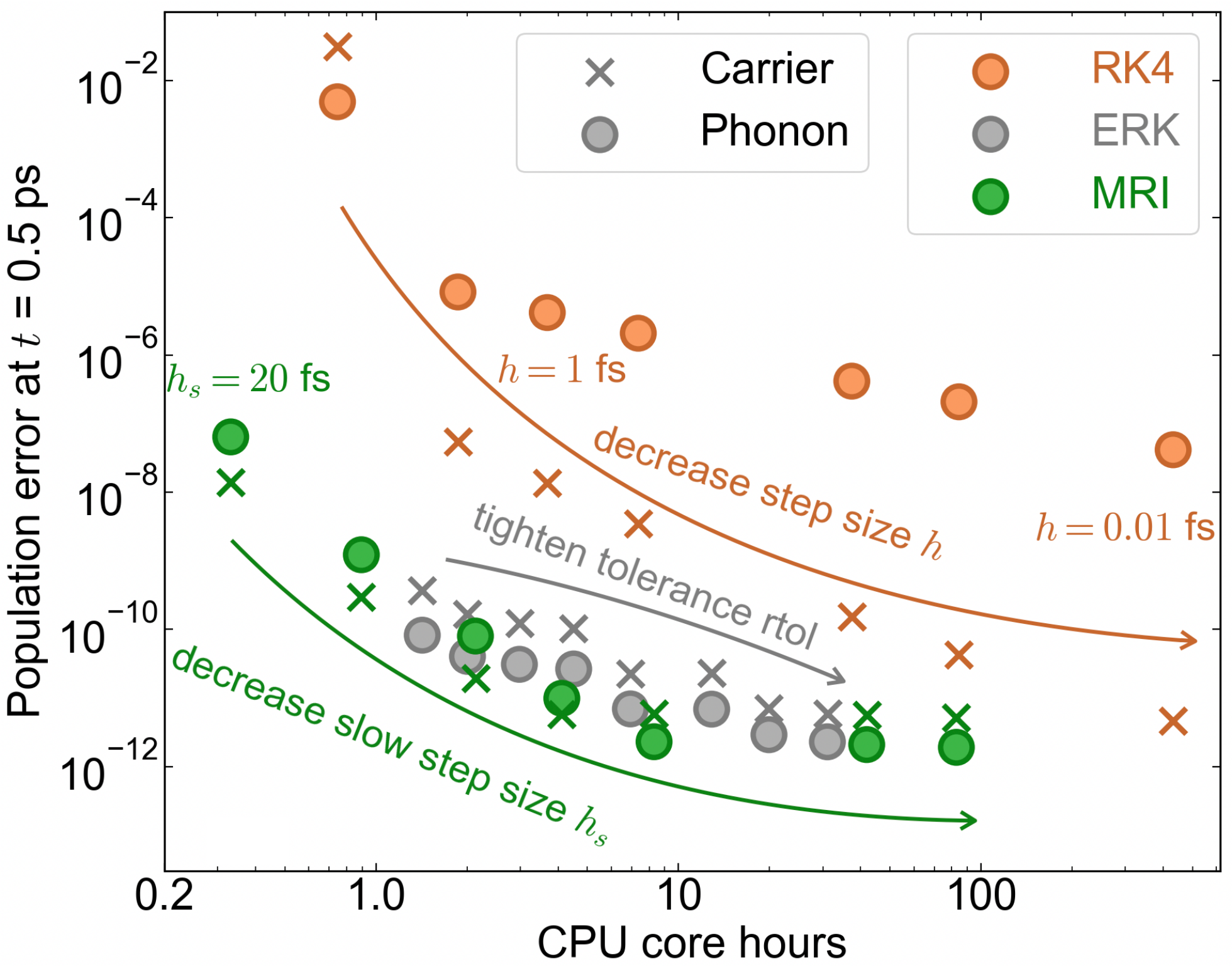}
\caption{\textbf{Comparison of error and cost for solving the rt-BTE with different methods.} The carrier and phonon population errors at $t = 0.5$~ps as a function of CPU core-hour cost, shown for RK4 (orange), ERK (gray), and MRI (green) time integration methods.  
For RK4, points from left to right represent results using progressively smaller time steps, $h$, from 5 fs to 0.01 fs. For MRI, results from left to right correspond to slow time steps, $h_s$, ranging from 20 fs to 0.05 fs. For ERK, the relative tolerance ranges from $10^{-4}$ to $10^{-11}$ from left to right with a fixed absolute tolerance. These parameter changes are indicated schematically with arrows for each method. 
See Methods for detailed tolerance settings.}
\label{fig2}
\end{figure}

%%% Describe the purpose of figure 2
Figure~\ref{fig2} provides a detailed benchmark of the efficiency improvements obtained with the adaptive and multirate schemes for these graphene simulations. We compare results obtained using RK4 with a fixed time step, an adaptive 4th-order ERK method, and a third-order MRI method.
% by simulating the dynamics 
%%% Introduce the parameters that affect solution error.
The accuracy and computational cost depend strongly on the choice of the following key parameters: the fixed time step size $h$ in RK4, the fixed slow time step size $h_s$ in the MRI method, and the tolerances in the standalone ERK method (and also in ERK used for the MRI fast timescale). 
%%% Describe the x and y axes of the plot
In Fig.~\ref{fig2}, we vary these parameters for each method and plot the carrier and phonon population errors at $t=0.5$~ps against the corresponding computational cost (runtime). 
%%% Describe the definition of error
The error is taken relative to a reference MRI solution with a small step size $h_s$; see Methods for details.

%%% Describe figure - RK4
With RK4, the solution error can be reduced systematically by using smaller time steps at the expense of taking longer computation time. When the time step is too long (here $h>5$~fs), the populations diverge at long enough simulation times. 
%%% Describe figure - ERK
Error-versus-cost results with the ERK method are obtained by tightening the relative tolerance (rtol) from $10^{-4}$ to $10^{-11}$ while keeping the absolute tolerance fixed.
%%% Describe figure - MRI
Results using the MRI method are obtained by varying $h_s$ with fixed tolerance values for the adaptive ERK method used for the fast timescale.

\begin{figure}
\centering
\includegraphics[width=0.95\columnwidth]{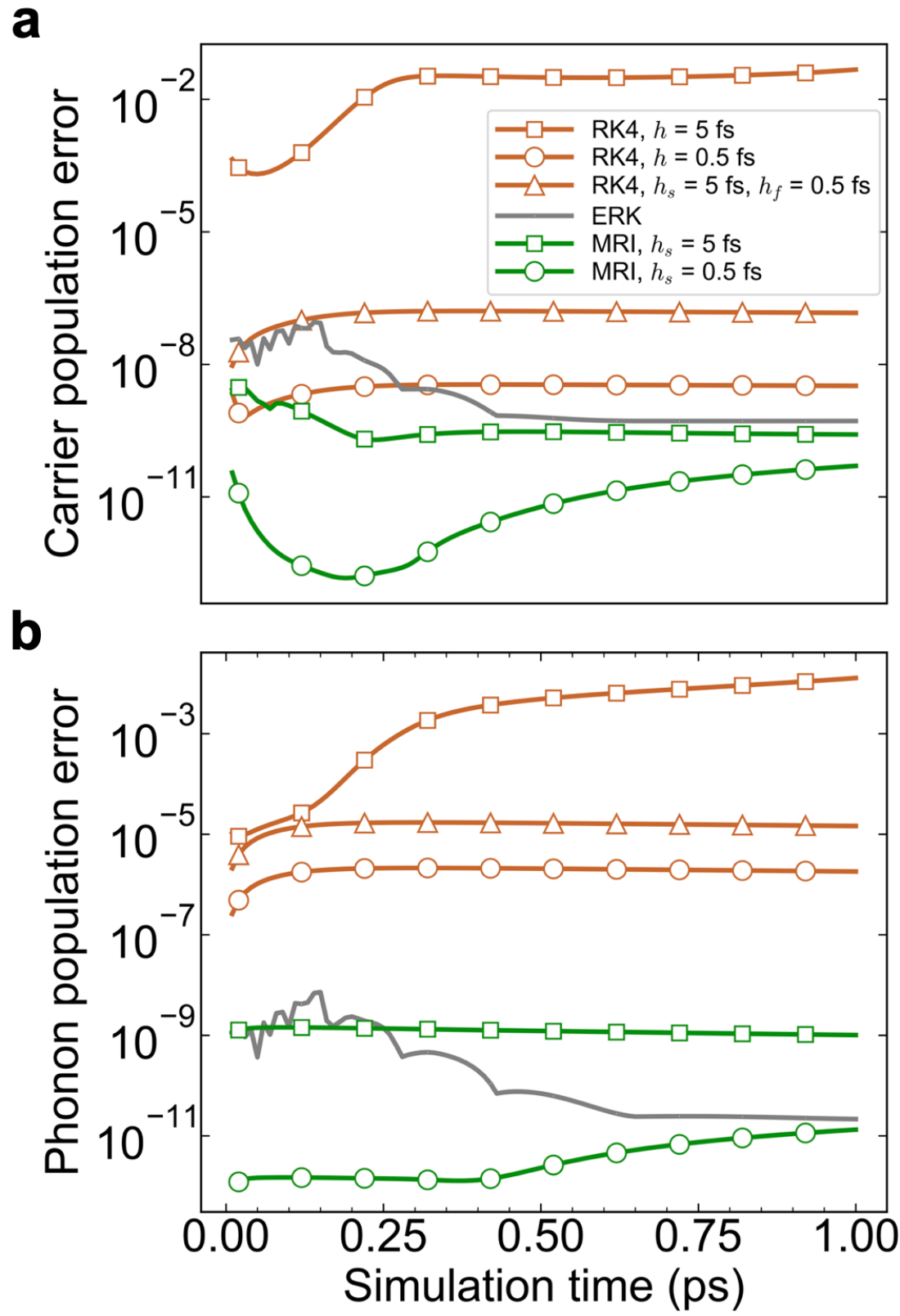}
\caption{\textbf{Solution error for the coupled dynamics in graphene.} Population errors for \textbf{a} carrier and \textbf{b} phonon dynamics as a function of time using different time integration methods and parameters. The reference solution is the same as in Fig.~\ref{fig2}. Orange curves show solution errors for RK4 with fixed time steps $h = 5$~fs (square) and $0.5$~fs (circle), or an operator splitting approach with RK4 using separate time steps for phonons (5~fs) and electrons (0.5~fs) (triangle). The green curves show results for MRI with $h_s = 5$~fs (square) and $0.5$~fs (circle). The gray curve shows results for ERK with relative tolerance $10^{-5}$ and same abolute tolerance as in Fig.~\ref{fig2}. See Methods for detailed tolerance settings.}
\label{fig3}
\end{figure}

\begin{figure*}
\centering
\includegraphics[width=1.0\textwidth]{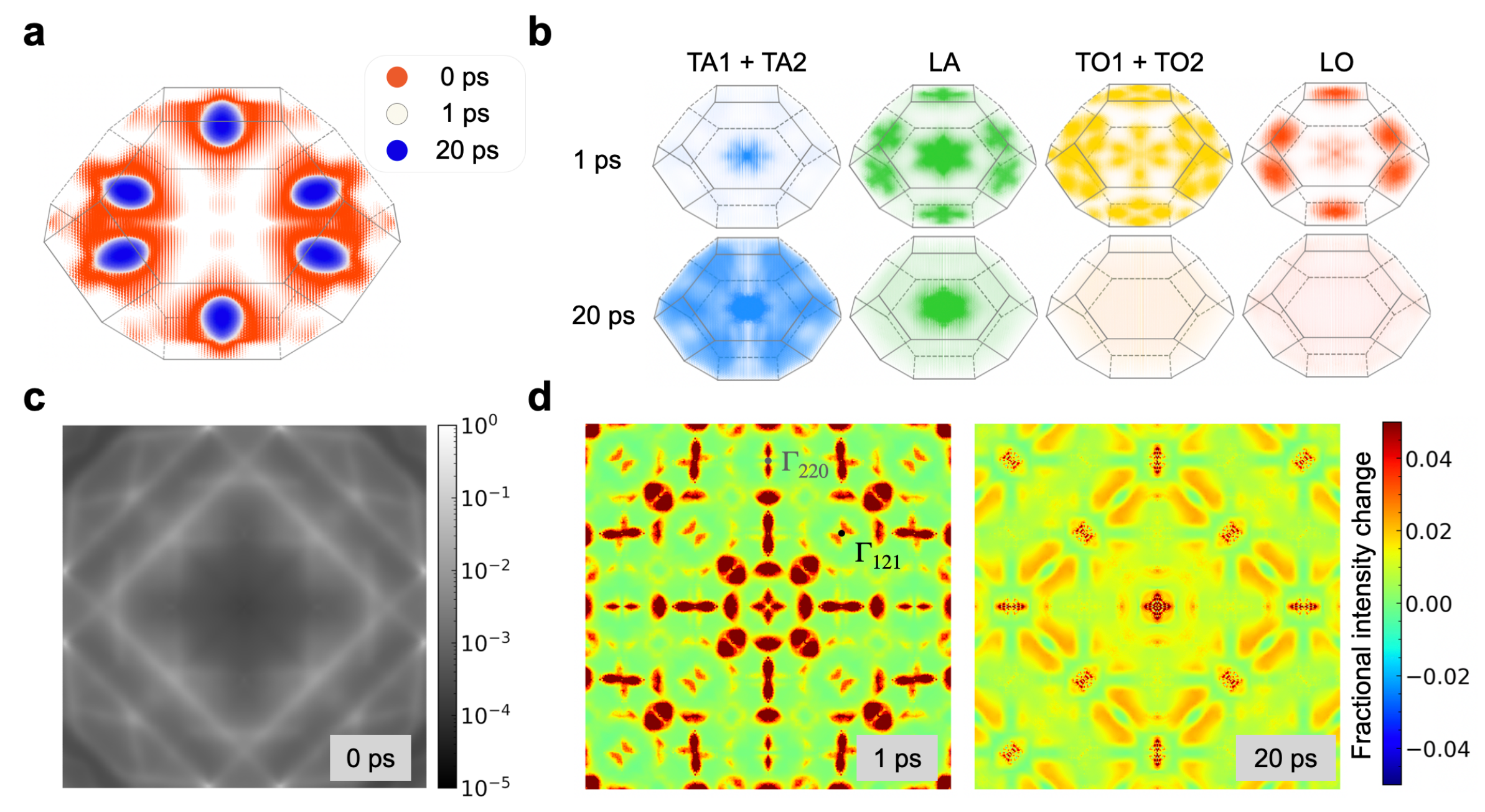}
\caption{\textbf{Coupled dynamics of electrons and phonons in silicon.} \textbf{a} Electron population, $f_{n\mathbf{k}}(t)$, in the first BZ at three times up to 20 ps. \textbf{b} Change in phonon population, $\Delta N_{\nu \mathbf{q}}(t)$, in the first BZ at times 1~ps and 20~ps, plotted separately for transverse acoustic (TA1 + TA2), longitudinal acoustic (LA), transverse optical (TO1 + TO2), and longitudinal optical (LO) modes. \textbf{c} Simulated TDS intensity, $I(\mathbf{q},t = 0)$ in thermal equilibrium at 300 K, for (100) crystal orientation with 28 keV incident X-ray. The intensity is shown in log-scale, normalized by the maximum intensity. \textbf{d} Fractional change in diffuse scattering intensity, $\Delta I(\mathbf{q},t)$, for (100) orientation at 1 ps and 20 ps simulation times.}
\label{fig4}
\end{figure*}
%

%%% Where do the results lie, overall conclusion
Results for the ERK and MRI methods are in the lower-left corner of the error-versus-cost map in Fig.~\ref{fig2}, which shows their superior performance compared to RK4 with a fixed time step. 
These methods are more efficient over a wide range of tolerances (ERK) and slow time step values (MRI), and are particularly robust in reducing phonon population errors. 

%%% How does MRI compare to RK4 for both carrier and phonon
Detailed examples are helpful to quantify the performance improvement. 
The carrier population error using the MRI method with $h_s = 20$~fs is comparable to RK4 with a much shorter time step ($h = 1$~fs, a typical step size chosen for $e$-ph dynamics~\cite{Tong-BTE2021}), but the MRI method requires only $\sim$10\% of the computational effort of RK4 for a 0.5-ps simulation.
In addition, the phonon population error with the MRI method for the same settings is comparable to RK4 with an extremely short time step, $h = 0.01$~fs, but the computational cost for phonon dynamics is $1000$x smaller relative to RK4. %\kelly{Marco says he will edit this part further.}
%%% comparing the same cost
Analyzed differently, for the same computational cost as RK4 with $h = 1$~fs, the MRI method improves the carrier accuracy by \mbox{3 orders} of magnitude and the phonon accuracy by 6 orders of magnitude.
This dramatic improvement of efficiency and accuracy in the MRI method is game-changing---for example, it enables modeling nonequilibrium lattice dynamics up to long times ($> 100$~ps) and in bulk materials, as we show below. 
%%% Additional details 
Additional benchmarks for the ERK and MRI methods are shown in \mbox{Supplementary} Fig.~1-2.

%%% Describe figure 3 and tolerance settings
The carrier and phonon population errors during the simulated electron and phonon dynamics are shown in Fig.~\ref{fig3} for different methods and time-stepping parameters. 
%%% Overall error behavior and how it relates to figure 2
In all simulations, we find that the error approaches a constant value after 0.5 ps, showing that the error-cost results in Fig.~\ref{fig2} are representative of the entire dynamics up to long simulation times. 
%%% Compare MRI and RK4 with similar cost
With both the RK4 and MRI methods, the respective step sizes $h$ and $h_s$ determine how frequently the ph-ph collision integral $\mathcal{I}^{\mathrm{ph\text{-}ph}}$ is evaluated, which is the main computational cost driver. 
To make a fair comparison, we set $h = h_s$ in both \mbox{methods.} 
%to obtain equal computational effort in both methods. 
As shown in Fig.~\ref{fig3}, with this setup the MRI results are significantly more accurate than RK4, with accuracy greater by over 3 orders of magnitude for $h = 0.5$~fs and 8 orders of magnitude for $h = 5$~fs.

%%% RK4 with separate fixed time steps not as good as MRI
To highlight the superior performance of the MRI method, we also consider a first-order operator-splitting method where RK4 is employed with different fixed time steps for the fast ($e$-ph) and slow (ph-ph) components defined in Eq.~\eqref{eq:mri_ode}. In this case, one simulation step consists of an RK4 advance of the fast and slow parts with separate fixed time steps. %using $f_s$ with a step size of $h_s$ and an RK4 advance using $f_f$ with a step size of $h_f$
The results for this method with a 0.5~fs time step for $e$-ph scattering and 5~fs for ph-ph scattering are plotted with orange triangles in Fig.~\ref{fig3}. In terms of both speed and accuracy, this operator-splitting method falls between the RK4 results with $h = 0.5$~fs and $h = 5$~fs, without offering a clear improvement over RK4 with a single time step. This time-splitting approach is therefore significantly less efficient and accurate than the MRI and ERK methods. The same trends hold for other combinations of fixed fast and slow time steps when using the operator splitting method with RK4.

%%% conclusion for graphene
Our results for graphene demonstrate that the ERK and MRI methods can efficiently solve the coupled rt-BTE by taking advantage of the inherently different timescales of electron and phonon dynamics. 
More broadly, these findings suggest that adaptive time-stepping can improve the efficiency and accuracy of simulations involving coupled degrees of freedom in materials. 

\subsection{Electron and phonon dynamics in silicon}

We address the challenge of simulating lattice dynamics in bulk materials using first-principles $e$-ph and ph-ph scattering processes on dense momentum grids, choosing silicon (Si) as a case study. 
This calculation is not feasible with fixed-step time integration methods such as RK4, even with the high-performance rt-BTE parallel implementation in \textsc{Perturbo}~\cite{Zhou-perturbo2021}. 
As such, we employ a third-order MRI method paired with an adaptive ERK method as the fast timescale integrator for this simulation.
At time zero, we initialize excited electrons in the conduction band of Si using a hot Fermi-Dirac distribution at 2000 K (with concentration $4.6\times10^{20}\,\mathrm{cm}^{-3}$) while phonons are initially in thermal equilibrium at \mbox{300 K.}

The coupled electron and phonon dynamics in Si are \mbox{analyzed} in Fig.~\ref{fig4}. The relaxation of excited electrons to the six lowest conduction-band valleys is shown in Fig.~\ref{fig4}a and
the change in phonon populations for different classes of phonon modes is visualized in Fig.~\ref{fig4}b.
In the first ps, most phonons are excited in the optical and longitudinal acoustic modes via $e$-ph interactions, mainly near the edge of the BZ or along the $\Gamma$-$X$ high-symmetry line.
Between 1 ps and 20 ps, these optically excited phonons decay into acoustic phonons, which progressively thermalize to long-wavelength lattice vibrations with $q\rightarrow0$ ($\Gamma$ point in Fig.~\ref{fig4}b).

Using the time-dependent phonon populations, we can connect our results with widely used experimental probes of time-domain lattice dynamics, particularly ultrafast diffraction techniques.
Momentum-resolved thermal diffuse X-ray scattering (TDS) has been used extensively to determine phonon dispersions and study ultrafast phonon dynamics~\cite{Holt-tds1999,Trigo-xray2013,Filippetto-TDSreview2022}. The transient TDS at scattering vector $\mathbf{q}$ at time $t$, $I(\mathbf{q},t)$, is obtained from the time-dependent phonon populations, phonon frequencies, and structure factor $F_{\nu}\,(\mathbf{q},t)$, as described in Ref.~\cite{Warren-xray1990} and Methods.

Figure~\ref{fig4}c shows the simulated TDS in Si by computing $I(\mathbf{q},t=0)$, with phonon populations at thermal equilibrium at 300 K.
Each point on the plot corresponds to a scattering vector $\mathbf{q}$ on the Ewald sphere cutting through reciprocal space for an incident X-ray energy of 28~keV. 
The results agree with both experimental measurements and theoretical predictions in Ref.~\cite{Holt-tds1999}. 
Fig.~\ref{fig4}d shows the relative intensity change $\Delta I(\mathbf{q},t) = \frac{I(\mathbf{q},t)-I(\mathbf{q},t=0)}{I(\mathbf{q},t=0)}$ at 1 ps and 20 ps on the (100) plane (without Ewald-sphere projection), where 
$\Gamma_{000}$ is at the center of the plot and $\Gamma_{220}$ and $\Gamma_{121}$ are labeled in the plot. 
At 1 ps, the TDS primarily shows an increase in optical phonons from electron cooling. At 20 ps, the main contribution to the TDS is from acoustic phonons, with the LA modes \mbox{giving} the signal close to $\Gamma$ and the TA modes the \textquotedblleft double-bar\textquotedblright~patterns.

These results show that adaptive and multirate time-stepping of the rt-BTEs enables first-principles simulations of phonon dynamics up to long timescales in bulk crystals, as well as modeling TDS experiments that are widely used to probe ultrafast lattice dynamics.

\section{Discussion}\label{s:discussion}
Time step adaptivity is a key factor in improving the accuracy and efficiency of rt-BTE simulations. Therefore, it is instructive to analyze the time step sizes selected by the algorithm over the course of the simulation as the dynamics evolve. At early stages of the carrier dynamics in the graphene simulation, the adaptive step size in the ERK method is close to zero, and then it increases to about $5$~fs after 400-fs simulation time (see Supplementary Fig.~1d). Compared to the RK4 method, ERK provides significantly more accurate results for both carriers and phonons when the parameters are chosen to obtain the same computational cost in both methods. This accuracy underscores the need to resolve early-time dynamics with smaller time steps to minimize error propagation to longer times. With reasonable tolerances and slow time-step size choices, the average adaptive step with the ERK method and the fast time step in the MRI method both reach a steady state value of 5 fs, suggesting that adaptive methods can find the characteristic timescales of physical interactions in the system.

In addition to improving efficiency, using adaptive methods eliminates the need to converge the solution with respect to the chosen time step. In adaptive ERK methods, the only external parameter is the desired level of accuracy for the solution, while in the MRI method the slow time step is estimated using the ph-ph coupling strength. Both methods will produce greater efficiency than fixed time step methods with these settings without any fine-tuning of simulation parameters.
(The only point to note is that if the chosen slow time-step size is too small, the solution will be more accurate than required, as the adaptive fast time step is constrained to be smaller than the slow time step, leading to an unnecessarily high computation cost.) 
As \textsc{sundials} continues to expand the capabilities of multirate methods, in future work it will become possible to further improve the efficiency by having the algorithm select both the fast and slow time steps based on user-defined tolerances.

The range of nonequilibrium dynamics that can be addressed with multirate methods is not limited to the coupled electron and phonon system studied here.
For example, the nonequilibrium dynamics of excitons and other elementary excitations can also be studied from first principles, including their couplings with the lattice~\cite{Chen-exciton2020, Chen-exciton2022, Perfetto-exciton2024}.
Several frameworks beyond the rt-BTE, including time-dependent density functional theory (DFT) with (non)adiabatic lattice dynamics, master-equation methods such as the  Liouville or Lindblad equations~\cite{Rosati-lindblad2014}, and nonequilibrium Green's functions could all greatly benefit from the multirate and adaptive time-stepping shown in this work. This advance provides a new tool for addressing a potentially wide range of problems in nonequilibrium materials physics, including modeling and interpreting time-domain spectroscopy and diffraction experiments.

In summary, we applied adaptive and multirate time-stepping methods to accelerate simulations of coupled electron and phonon dynamics using the rt-BTE. We achieve a 10- to 100-fold speed-up relative to fixed time step method with the same accuracy, or orders of magnitude greater accuracy when setting simulation parameters to enforce equal computational cost. This enables affordable simulations of time-domain lattice dynamics in both 2D and bulk materials.
Our research lays the groundwork for studying coupled electron and phonon dynamics in a broader array of materials. Future directions include adding explicit treatments of light pulses, coherent electron and phonon dynamics, and higher-order phonon interactions to explore novel quantum states and dynamical control of materials, such as phonon-driven Floquet engineering and time-domain tuning of physical properties, order parameters, and crystal structures. 

\section{Methods}\label{s:methods}

\subsection{Electron-phonon and phonon-phonon scattering from first principles}\label{methods:A}
%\vspace{-10pt}
%%% From DFT to IFC
We carry out first-principles calculations of $e$-ph and ph-ph scattering in graphene and silicon. We obtain the electronic ground state and band structure from density functional theory (DFT)~\cite{Martin-DFT2020} with the \textsc{Quantum ESPRESSO}~\cite{QE2017} package, using the local density approximation and norm-conserving pseudopotentials.

%%% e-ph matrix elements
The $e$-ph matrix elements are defined as~\cite{Zhou-perturbo2021} 
\begin{equation}
g_{mn\nu}\,(\mathbf{k},\mathbf{q}) = \sqrt{\frac{\hbar}{2\omega_{\nu\textbf{q}}}}
\sum_{\kappa \alpha} \frac{e_{\nu\mathbf{q}}^{\kappa\alpha}}{\sqrt{
\mu_\kappa}}
\langle \psi_{m\textbf{k}+\textbf{q}}\lvert 
\partial_{\textbf{q},\kappa\alpha} V
\rvert \psi_{n\textbf{k}}\rangle
\end{equation}
%%% describe 
and represent the probability amplitude to scatter from an initial electronic state $\rvert \psi_{n\textbf{k}}\rangle$, with band index $n$, crystal momentum $\textbf{k}$, and energy $\varepsilon_{n\textbf{k}}$, to a final state $\rvert \psi_{m\textbf{k}+\textbf{q}}\rangle$ by emitting or absorbing a phonon with mode index $\nu$, wave-vector $\textbf{q}$, frequency $\omega_{\nu\textbf{q}}$, and displacement eigenvector $\textbf{e}_{\nu\textbf{q}}$ due to the $e$-ph perturbation potential $\partial_{\textbf{q},\kappa\alpha} V$, where $\mu_\kappa$ is the mass of atom $\kappa$ in the unit cell and $\alpha$ labels Cartesian components. 
%%% how we get the bra-ket
The matrix elements are first computed on a coarse momentum grid using density functional perturbation theory (DFPT)~\cite{Baroni-DFT2001}. In the \textsc{Perturbo} code, the $e$-ph matrix elements are then transformed to a maximally-localized Wannier basis (generated using \textsc{wannier90}~\cite{Mostofi-wannier2014}) and interpolated to finer electron and phonon momentum grids.

For the rt-BTE dynamics, the $e$-ph and ph-$e$ scattering integrals at time $t$ read~\cite{mahan-condensed2010,Tong-BTE2021}
%%% scatterting integral $e$-ph
\begin{equation}
\label{eq:e-ph}
    \begin{split}
    % \mathcal{I}^eph
        \mathcal{I}^{e\text{-}\mathrm{ph}}[f_{n\mathbf{k}},N_{\nu\mathbf{q}}]
        % eph constant
        = &- \frac{2\pi}{\hbar} \frac{1}{\mathcal{N}_\textbf{q}}
        % gmatrix
        \sum_{m\nu\textbf{q}} \lvert g_{mn\nu} (\textbf{k}, \textbf{q})\rvert^2 \\ 
        % start of braket
        &\times \Big[
        % femission
        F_{\rm{em}}\times \delta(\varepsilon_{n\textbf{k}} -\hbar \omega_{\nu\textbf{q}} -\varepsilon_{m\textbf{k}+\textbf{q}})\\
        % fabsorption
        &+F_{\rm{abs}}\times \delta(\varepsilon_{n\textbf{k}} +\hbar \omega_{\nu\textbf{q}} -\varepsilon_{m\textbf{k}+\textbf{q}})\Big],\\
% I ph-e
\mathcal{I}^{\mathrm{ph}\text{-}e}[f_{n\mathbf{k}},N_{\nu\mathbf{q}}]
% eph constant
=&- \frac{4\pi}{\hbar} \frac{1}{\mathcal{N}_\textbf{k}}
% gmatrix
\sum_{mn\textbf{k}} \lvert g_{mn\nu} (\textbf{k}, \textbf{q})\rvert^2 \\        
% fabsoprtion
&\times F_{\rm{abs}}\times \delta(\varepsilon_{n\textbf{k}} +\hbar \omega_{\nu\textbf{q}} -\varepsilon_{m\textbf{k}+\textbf{q}}),
\end{split}
\end{equation}
where $\mathcal{N}_\textbf{k}$ and $\mathcal{N}_\textbf{q}$ are the total number of momentum grid points for electrons and phonons, respectively. In practice, a subset of the $\mathcal{N}_\textbf{k}$ electron momenta within a given energy range of the band edge or Fermi energy is included in the dynamics (and in the $e$-ph scattering integral summations) to reduce computational cost. The emission and absorption factors used above are defined as~\cite{Tong-BTE2021}
% explain emmesion and absorption
\begin{equation}
\begin{split}
F_{\rm{em}} &= f_{n\textbf{k}} (1-f_{m\textbf{k}+\textbf{q}})(1+N_{\nu\textbf{q}}) \\
 & \quad -(1-f_{n\textbf{k}}) f_{m\textbf{k}+\textbf{q}} N_{\nu\textbf{q}},
\end{split}
\end{equation}
and
\begin{equation}
\begin{split}
    F_{\rm{abs}} &= f_{n\textbf{k}} (1-f_{m\textbf{k}+\textbf{q}})N_{\nu\textbf{q}}\\
    & \quad -(1-f_{n\textbf{k}}) f_{m\textbf{k}+\textbf{q}} (1+N_{\nu\textbf{q}}).
\end{split}
\end{equation}
The delta functions enforcing energy conservation in Eq.~\ref{eq:e-ph} are approximated by the Gaussian function $\delta(E) \approx\frac{\pi}{\sigma} \exp(E/\sigma)^2$, where $\sigma$ is a broadening parameter typically set to a few meV.

%%% IFC definition
We compute the second- and third-order inter-atomic force constants (IFCs), $\Phi_{\alpha\beta}(l\kappa;l'\kappa')$ and $\Psi_{\alpha \beta\gamma}(l\kappa;l'\kappa';l'' \kappa'')$ respectively, by expanding the potential energy of the system with respect to small atomic displacements $\textbf{u}$:
\begin{equation}
\begin{split}
        U &= U_0 + \frac{1}{2!} \sum_{ll'}\sum_{\kappa\kappa'}\sum_{\alpha\beta}
        \Phi_{\alpha\beta}(l\kappa;l'\kappa')
        u_{l\kappa}^{\alpha} u_{l'\kappa'}^{\beta}\\
        & + \frac{1}{3!} \sum_{ll'l''}\sum_{\kappa\kappa'\kappa''}\sum_{\alpha\beta\gamma}
        \Psi_{\alpha\beta\gamma}(l\kappa;l'\kappa';l''\kappa'')
        u_{l\kappa}^{\alpha} u_{l'\kappa'}^{\beta}u_{l''\kappa''}^{\gamma}.
\end{split}
\end{equation}
Here, $U_0$ is the total potential energy for atoms in equilibrium, and the vector $\textbf{u}_{l\kappa}$ is the displacement from equilibrium of atom $\kappa$ in the $l$-th unit cell, while $\alpha$, $\beta$, and $\gamma$ label Cartesian coordinates. 
%%% TDEP configuration
We compute the IFCs using the temperature-dependent effective poential (TDEP) method~\cite{Hellman-tdep2013}, where the IFCs are obtained by computing the atomic forces on structures with random thermal displacements distributed according to a canonical ensemble at 300 K. These calculations employ supercells with dimensions $12\times12\times1$ for graphene and $6\times 6\times 6$ for silicon.
The second-order IFCs are used to compute the phonon frequencies and eigenvectors~\cite{Zhou-STO2018}, and the third-order IFCs to compute the ph-ph matrix elements. 

The ph-ph matrix elements describe the interaction of three phonons with momenta $\textbf{q}$, $\textbf{q}'$, and $\textbf{q}''$ and mode indices $\nu$, $\nu'$, and $\nu''$. They are computed as Fourier transforms of the third-order IFCs using
\begin{equation}
\begin{split}
%phi
\Psi_{\nu\nu'\nu''}\,(\mathbf{q},\mathbf{q'},\mathbf{q''}) 
% constant
&= \frac{1}{6} \sqrt{\frac{\hbar^3}{8
\omega_{\nu\mathbf{q}}
\omega_{\nu'\mathbf{q'}}
\omega_{\nu''\mathbf{q''}}}}
% delta
\Delta(\mathbf{q}+\mathbf{q}'+\mathbf{q}'')\\
% sums
&\times \sum_{l' l''} % exp(ikr)
\sum_{\kappa \kappa'\kappa''} \sum_{\alpha\beta\gamma}
\exp{\big[i (\mathbf{q'}\cdot \mathbf{R}_{l'} + \mathbf{q''}\cdot \mathbf{R}_{ l''})\big]}\\
&\frac{% eigenstates
e_{\nu\mathbf{q}}^{\kappa\alpha}
e_{\nu'\mathbf{q}'}^{\kappa'\beta}
e_{\nu'' \mathbf{q}''}^{\kappa''\gamma}}
{\sqrt{\mu_{\kappa}\mu_{\kappa'}\mu_{\kappa''}}}
% ifc
\Psi_{\alpha \beta\gamma}(0\kappa;l'\kappa';l'' \kappa'')
\end{split}
\end{equation}
where $\mathbf{R}_l$ is the lattice vector of the unit cell $l$, and the delta function $\Delta(\mathbf{q}+\mathbf{q}'+\mathbf{q}'')$ conserves crystal momentum, ensuring $\mathbf{q}+\mathbf{q}'+\mathbf{q}''$ is equal to zero or a reciprocal lattice vector. 

The ph-ph scattering integral can be written as
\begin{equation}\label{eq:ph-ph}
\begin{split}
\mathcal{I}^{\mathrm{ph\text{-}ph}}&[N_{\nu\mathbf{q}}]
% eph constant
=-\frac{36\pi}{\hbar} \frac{1}{\mathcal{N}_\textbf{q}}
% sum
\sum_{\nu'\nu''}\sum_{\textbf{q}'\textbf{q}''}
% phimatrix
\lvert \Psi_{\nu\nu'\nu''} (\textbf{q}, \textbf{q}',\textbf{q}'')\rvert^2\\
% times
&\times\Big[
% delta 1
G_{\rm{em}} \times \delta{(-\hbar\omega_{\nu\textbf{q}}+\hbar\omega_{\nu'\textbf{q}'}+\hbar\omega_{\nu''\textbf{q}''})} \\
% delta 2
&+ 2 \, G_{\rm{abs}} \times \delta{(\hbar\omega_{\nu\textbf{q}}-\hbar\omega_{\nu'\textbf{q}'}+\hbar\omega_{\nu''\textbf{q}''})}
\Big],
\end{split}
\end{equation}
where
\begin{equation}
\begin{split}
    G_{\rm{em}} &= N_{\nu\textbf{q}}(N_{\nu'\textbf{q}'}+1)(N_{\nu''\textbf{q}''}+1)\\
    & \quad - (N_{\nu\textbf{q}}+1)N_{\nu'\textbf{q}'}N_{\nu''\textbf{q}''} ,
\end{split}
\end{equation}
and
\begin{equation}
\begin{split}
    G_{\rm{abs}} &= N_{\nu\textbf{q}}(N_{\nu'\textbf{q}'}+1)N_{\nu''\textbf{q}''} \\
    & \quad - (N_{\nu\textbf{q}}+1)N_{\nu'\textbf{q}'}(N_{\nu''\textbf{q}''}+1).
\end{split}
\end{equation}
The Wannier interpolation, $e$-ph and ph-ph matrix computation, and ultrafast dynamics simulation described above are implemented in the \textsc{Perturbo} package~\cite{Zhou-perturbo2021}. Both MPI and OpenMP parallelization are employed to sum over scattering channels when computing $e$-ph and ph-ph matrices, and to sum over $\mathbf{k}$- and $\mathbf{q}$-points when computing the scattering integrals in Eq.~\eqref{eq:e-ph} and Eq.~\eqref{eq:ph-ph}. 

%%% explain the grid sizes - graphene
For graphene, the phonon scattering rates are converged for ph-ph matrix elements computed on a $\textbf{q}$-grid with size $200\times200\times1$, while full convergence of $e$-ph scattering requires a $400\times400\times1$ $\textbf{q}$-grid. %To balance accuracy and memory usage, 
The nonequilibrium dynamics calculation in Fig.~\ref{fig1} employs this very dense $400\times400\times1$ grid, with ph-ph matrix elements interpolated on the fly. The efficiency analyses in Fig.~\ref{fig2}, Fig.~\ref{fig3} and Supplementary Figs.~1$-$3 are obtained using $200\times200\times1$ $\textbf{k}$- and $\textbf{q}$-grids. A small phonon frequency cutoff of 1 meV is employed. We use a Gaussian broadening of 20 meV for the collision integrals $\mathcal{I}^{e\text{-}\mathrm{ph}}$ and $\mathcal{I}^{\mathrm{ph}\text{-}e}$, and 2 meV for $\mathcal{I}^{\mathrm{ph\text{-}ph}}$.

%%% explain the grid sizes - silicon
For silicon, the ph-ph matrix elements are computed on a $40\times40\times40$ $\textbf{q}$-grid with a phonon frequency cutoff of 0.2 meV. The $e$-ph matrix elements and the dynamics are computed on a finer $80\times80\times80$ $\textbf{q}$-grid. During the dynamics calculations, the ph-ph matrix elements are interpolated onto this finer grid. The Gaussian broadening is 5 meV for $\mathcal{I}^{e\text{-}\mathrm{ph}}$ and $\mathcal{I}^{\mathrm{ph}\text{-}e}$ and 1.1 meV for $\mathcal{I}^{\mathrm{ph\text{-}ph}}$. To find optimal grids for the nonequilibrium lattice dynamics, we compute mode-resolved three-phonon scattering rates at 300~K with the same ph-ph matrix elements employed in the time-domain simulations (see Supplementary Fig.~4). The coupled dynamics employ the default MRI method in \textsc{arkode} with $h_s = 50$~fs. The fast timescale is integrated with the default adaptive ERK method in \textsc{arkode} using a relative tolerance $\mathrm{rtol} = 10^{-5}$ and absolute tolerances $\mathrm{atol}_c = 10^{-9}$ and $\mathrm{atol_{ph}} = 10^{-11}$ for carriers and phonons, respectively. Convergence at these tolerance values is confirmed by comparing with tests using tighter tolerances and smaller $h_s$ values. An analysis of the effective phonon temperatures is provided in Supplementary Fig.~5 to elucidate the thermalization process of each phonon mode. 

%%% TDS
The transient TDS at scattering vector $\mathbf{q}$ and time $t$, $I(\mathbf{q},t)$, is given by 
\begin{equation}
    I(\mathbf{q},t) \propto \sum_{j} \frac{1}{\omega_{\nu\mathbf{q'}}} \Big[ N_{\nu\mathbf{q'}}(t)+\frac{1}{2} \Big] \lvert F_{\nu}\,(\mathbf{q},t) \rvert^2, 
\end{equation}
where $\mathbf{q'} = \mathbf{q} - \mathbf{K}_{\mathbf{q}}$ is the momentum $\mathbf{q}$ folded to the first BZ, and $\mathbf{K}_{\mathbf{q}}$ is the nearest reciprocal lattice vector to $\mathbf{q}$. The one-phonon structure factor $F_{\nu}\,(\mathbf{q},t)$ is defined as
\begin{equation}
    F_{\nu}\,(\mathbf{q},t) = \sum_{\kappa} \frac{f_{\kappa}(\mathbf{q})}{\sqrt{\mu_{\kappa}}} \exp \, [-M_{\kappa}(\mathbf{q})] (\mathbf{q} \cdot \mathbf{e}^{\kappa}_{\nu\mathbf{q'}} )\,\exp \, (-i \mathbf{K}_{\mathbf{q}} \cdot \mathbf{r}_{\kappa}) 
\end{equation}
where $f_{\kappa}(\mathbf{q})$ is the atomic scattering factor~\cite{Olukayode-xray2023}, $\mathbf{r}_{\kappa}$ is the equilibrium position of atom $\kappa$ in the unit cell, and $M_{\kappa}(\mathbf{q})$ is the Debye-Waller factor.

\subsection{Adaptive integrators and implementation}\label{methods:B}

\textbf{ERK method.}
We describe how a single step is carried out with an adaptive ERK method. For Eq.~\eqref{eq:erk_ode}, to advance from $t_{k-1}$ to $t_k = t_{k-1} + h_k$ with a method of order $p$, we use 
\begin{subequations}
  \label{eq:ERK_method}
  \begin{align}
    \label{eq:ERK_stage}
    z_i &= y_{k-1} + h_k \sum_{j=1}^{i-1} A_{i,j} f(t_{k,j}, z_j),
      \, i=1,\ldots,S, \\
    y_k &= y_{k-1} + h_k \sum_{i=1}^{S} b_i f(t_{k,i}, z_i), \\
    \tilde{y}_k &= y_{k-1} + h_k \sum_{i=1}^{S} \tilde{b}_i f(t_{k,i}, z_i),
  \end{align}
\end{subequations}
where there are $S$ stages, $z_i$, at times $t_{k,j} = t_{k-1} + c_j h_k$ and the coefficients that define the method for obtaining the new solution, $y_k$, are given in the corresponding Butcher tableau with $A \in \mathbb{R}^{S \times S}$, $b\in\mathbb{R}^S$, and $c\in\mathbb{R}^S$. The coefficients $\tilde{b}\in\mathbb{R}^S$ are used to obtain an embedded solution, $\tilde{y}_k$, (typically of order $p-1$) and the difference between the solution and embedding, $y_k - \tilde{y}_k$, provides an estimate of the local truncation error (LTE) for adapting the step size \cite{hairer2008solving}.
The ERK results above utilize the default adaptive ERK method in \textsc{arkode}, which is a five-stage, fourth-order method from Zonneveld \cite{Zonneveld1963automatic}. 

An attempted time step is accepted if it satisfies $\|{\rm{LTE}}\|_{\textsc{wrms}} \leq 1$ in the weighted root-mean-square (WRMS) norm,
\begin{equation} \label{eq:wrms_norm}
    \| v \|_{\textsc{wrms}} = \left( \frac{1}{\mathcal{N}_y} \sum_{i=1}^{\mathcal{N}_y} \left( v_i w_i \right)^2 \right)^{1/2},
\end{equation}
where $\mathcal{N}_{y}=\mathcal{N}_\mathrm{c}+\mathcal{N}_\mathrm{ph}$ is the length of the vector $v$ and the weights $w_i$ are defined by the most recent solution $y_{n-1}$, the relative tolerance (rtol) and the vector of absolute tolerances (atol) as 
\begin{equation} \label{eq:wrms_weights}
w_i =\big(\mathrm{rtol}\, \lvert y_{n-1,i} \rvert +\mathrm{atol_i} \big)^{-1}.
\end{equation}
The relative tolerance controls the number of digits of accuracy in the solution, while the absolute tolerance sets the level below which differences in small solution components are ignored.
If a step attempt is rejected, a new step size is computed based on the LTE, and the step is repeated. After successfully completing a step, the LTE is similarly used to determine the step size for the next step attempt.

In this work, we employ the default error controller in \textsc{arkode}, the PID controller \cite{soderlind1998automatic,soderlind2003digital,soderlind2006time}, for step size selection, 
\begin{equation} \label{eq:pid_controller}
h' = h_k 
\varepsilon_k^{-a_1/p_*}\,
\varepsilon_{k-1}^{-a_2/p_*}\,
\varepsilon_{k-2}^{-a_3/p_*},
\end{equation}
where $h'$ is the new step size and $\varepsilon_n$, $\varepsilon_{n-1}$, and $\varepsilon_{n-2}$ are the WRMS norms of the error estimates from the current and prior two time steps, respectively.
The values of $\varepsilon_{n-1}$ and $\varepsilon_{n-2}$ are initialized to $1$ and updated as steps are accepted.
We use the default PID parameter values ($a_1 = 0.58$, $a_2 = 0.21$, and $a_3 = 0.1$) except for $p_*$ where we use the method order, $p$, rather than the embedding order.
Additionally, we disable the step size adjustment thresholds, allowing any step size change between each step rather than retaining the current step size if the step growth factor is small. 
Previous work with \textsc{sundials} has shown that these adjustments to the default values are more efficient and lead to fewer failed steps, larger step sizes, more frequent changes in step size, and smoother step size profiles compared to the current default settings.

\textbf{MRI method.} 
Explicit MRI methods for Eq.~\eqref{eq:mri_ode} with $\hat{S}$ stages advance the solution from $t_{k-1}$ to $t_k = t_{k-1} + h_s$ with the following algorithm:
\begin{enumerate}
\item Set $z_1 = y_{k-1}$ and $\hat{t}_{k,1}=t_{k-1}$
\item For $i = 2,\ldots,\hat{S}$
\begin{enumerate}
\item Let $\hat{t}_{k,i} = t_{k-1} + \hat{c}_{i} h_s$ and $v(\hat{t}_{k,i-1}) = z_{i-1}$
\item \label{step:fastIVP} Solve $v'(t) = f_f(t, v) + r_i(t)$ on $t \in [\hat{t}_{k,i-1}, \hat{t}_{k,i}]$ where:
\begin{align*}
    r_i(t) &= \frac{1}{\Delta \hat{c}_i}\sum\limits_{j=1}^{i-1} \gamma_{i,j}\left(\tau\right) f^{s}(\hat{t}_{k,j}, z_j),\\
    \tau &= (t - \hat{t}_{k,i-1})/(\Delta \hat{c}_i h_s), \\
    \Delta \hat{c}_i &= \hat{c}_i - \hat{c}_{i-1}
\end{align*}
\item Set $z_i = v(\hat{t}_{k,i})$
\end{enumerate}
\item Set $y_{k} = z_{\hat{S}}$
\end{enumerate}
The abscissae are sorted, $0 = \hat{c}_1 \leq \cdots \leq \hat{c}_{\hat{S}} = 1$, and define the intervals over which the fast auxiliary ODE in step 2(b) is solved to compute the stage values $z_i$.
If $\Delta \hat{c}_i = 0$, the fast auxiliary ODE solve in step 2(b) reduces to a standard ERK stage update as in Eq.~\eqref{eq:ERK_method}.
The coefficient function,
\begin{equation}
  \gamma_{i,j}(\tau) = \sum_{k=1}^K \Omega_{i,j,k}\,\tau^{k-1},
\end{equation}
is a polynomial in time with coefficients $\Omega_{i,j,k} \in \mathbb{R}^{\hat{S} \times \hat{S} \times K}$ that define the coupling between slow and fast timescales.

The results above utilize the default MRI algorithm in \textsc{arkode}, a third-order multirate infinitesimal step (MIS) method \cite{schlegel2009multirate,schlegel2012implementation,schlegel2012numerical} based on the explicit method in Ref.~\cite{knoth1998implicit}. 
MIS methods are a subset of MRI methods where the coupling coefficients are uniquely defined from an ERK method with $S = \hat{S} - 1$ stages and sorted abscissae. 
For third-order accuracy, the ERK method must also satisfy an additional order condition \cite{knoth1998implicit}.
In the MIS case, the forcing function $r_i(t)$ reduces to a constant value ($K=1$) that depends on the stage index, $i$.
The resulting MIS method has abscissae $\hat{c} = [c_1\; \cdots\; c_{S}\; 1]^T$ and coupling coefficients:
\begin{equation}
  \label{eq:mis_coef}
  \Omega_{i,j,1} =
  \begin{cases}
    0, & \text{if}\; i=1,\\
    A_{i,j} - A_{i-1,j}, & \text{if}\; 2\le i\le S,\\
    b_{j} - A_{S-1,j}, & \text{if}\; i = S + 1 = \hat{S}
  \end{cases}.
\end{equation}

In theory, the auxiliary ODE in step 2(b) is solved exactly with an infinitesimally small step size; however, in practice, it is solved using any sufficiently accurate method. 
In this work, we use the same ERK method applied in the single-rate adaptive integration tests as the fast timescale integrator.
%
%\vspace{-10pt}
\subsection{Choices of tolerances and reference solutions}\label{methods:C}
%%% define error
%\vspace{-10pt}
Population errors in Fig.~\ref{fig2} and Fig.~\ref{fig3} are computed as $\lVert f_{n\textbf{k}}(t) - \tilde{f}_{n\textbf{k}}(t)\rVert /\lVert \tilde{f}_{n\textbf{k}}(t)\rVert$ for carriers and $ \lVert N_{\nu\textbf{q}}(t) - \tilde{N}_{\nu\textbf{q}}(t)\rVert / \lVert \tilde{N}_{\nu\textbf{q}}(t)\rVert$ for phonons.  In both figures, the reference solutions $\tilde{f}_{n\textbf{k}}(t)$ and $\tilde{N}_{\nu\textbf{q}}(t)$ are obtained using the MRI method with a very small slow time step, $h_s = 0.01$~fs, and tight tolerances in the fast timescale integration, $\mathrm{rtol} = 10^{-10}$ and $\mathrm{atol_c} = \mathrm{atol_{ph}} = 10^{-15}$ for both carriers and phonons.
%%% Define and justify reference solution
Analyzing solution errors with different reference solutions (Supplementary Fig.~S3) confirms the MRI method results converge significantly faster than RK4 across time step sizes. 
%%% Define and justify tolerance
In Fig.~\ref{fig2}, tolerance values for the MRI simulations (excluding the reference solution) range from $10^{-4}$ to $10^{-11}$ for the relative tolerance while the absolute tolerances are fixed as $\mathrm{atol_c} = 10^{-9}$ and $\mathrm{atol_{ph}} = 10^{-12}$ for carriers and phonons, respectively. For the ERK results, the relative tolerances also range from $10^{-4}$ to $10^{-11}$ and the absolute tolerances are fixed at $\mathrm{atol_c}$ = $\mathrm{atol_{ph}} = 10^{-15}$. 
In Fig.~\ref{fig3}, tolerances for both the ERK results and the fast timescale integration in the MRI method are set to $\mathrm{rtol}=10^{-5}$, $\mathrm{atol_c}=10^{-9}$, and $\mathrm{atol_{ph}}=10^{-12}$.
%
%\vspace{-10pt}
\subsection{Perturbo Interface with SUNDIALS}
\label{methods:D}
\vspace{-14pt}
%%% mention implementation of perturbo
This work required the development of a Fortran interface between \textsc{Perturbo} and \textsc{sundials}. While \textsc{sundials} is primarily written in C, the library provides Fortran interface modules generated by SWIG-Fortran \cite{Swig-Fortran} to facilitate interoperability with Fortran codes such as \textsc{Perturbo}.
For explicit methods, such as the ERK and MRI schemes considered in this work, interfacing with \textsc{sundials} requires providing the time integrator with a vector object to operate on state data, together with a single right-hand side function for an ERK method or fast and slow right-hand side functions for an MRI method. 
%\\
%\indent
To enable \textsc{sundials} to operate on \textsc{Perturbo} data, the 2D arrays for carrier and phonon populations are copied into a contiguous array under the native \textsc{sundials} serial vector.
The right-hand side functions provided to \textsc{sundials} are thin wrappers that unpack the input vector data into 2D Fortran arrays, compute the scattering integrals in parallel with the same subroutines utilized by the native RK4 method, and then pack result into the output \textsc{sundials} vector.
Overall, the implemented interface is versatile and efficient. For all calculations shown in this work, the CPU time spent on interfacing between \textsc{Perturbo} and \textsc{sundials} is negligible compared to the cost of the dynamics.\\
%
%\vspace{-10pt}
\begin{acknowledgments}
%\vspace{-10pt}
\noindent
This work is supported by the U.S. Department of Energy, Office of Science, Office of Advanced Scientific Computing Research and Office of Basic Energy Sciences, Scientific Discovery through Advanced Computing (SciDAC) program under Award Number DE-SC0022088. I.M. and M.B. acknowledge the support by the Liquid Sunlight Alliance, which is supported by the U.S. Department of Energy, Office of Science, Office of Basic Energy Sciences, under Award Number DE-SC0021266. For the development of the interface between \textsc{Perturbo} and \textsc{sundials}, K.Y. and M.B. were supported by the National Science Foundation under Grant No. OAC-2209262. This work was performed in part under the auspices of the U.S. Department of Energy by Lawrence Livermore National Laboratory under contract DE-AC52-07NA27344. LLNL-JRNL-2001089. This research used resources of the National Energy Research Scientific Computing Center (NERSC), a U.S. Department of Energy Office of Science User Facility located at Lawrence Berkeley National Laboratory, operated under Contract No. DE-AC02-05CH11231.
\end{acknowledgments}


\begin{thebibliography}{10}
\expandafter\ifx\csname url\endcsname\relax
  \def\url#1{\texttt{#1}}\fi
\expandafter\ifx\csname urlprefix\endcsname\relax\def\urlprefix{URL }\fi
\providecommand{\bibinfo}[2]{#2}
\providecommand{\eprint}[2][]{\url{#2}}

\bibitem{disa2021engineering}
\bibinfo{author}{Disa, A.~S.}, \bibinfo{author}{Nova, T.~F.} \& \bibinfo{author}{Cavalleri, A.}
\newblock \bibinfo{title}{Engineering crystal structures with light}.
\newblock \emph{\bibinfo{journal}{Nat. Phys.}} \textbf{\bibinfo{volume}{17}}, \bibinfo{pages}{1087--1092} (\bibinfo{year}{2021}).
\newblock \url{https://www.nature.com/articles/s41567-021-01366-1}.

\bibitem{basov2017towards}
\bibinfo{author}{Basov, D.}, \bibinfo{author}{Averitt, R.} \& \bibinfo{author}{Hsieh, D.}
\newblock \bibinfo{title}{Towards properties on demand in quantum materials}.
\newblock \emph{\bibinfo{journal}{Nat. Mater.}} \textbf{\bibinfo{volume}{16}}, \bibinfo{pages}{1077--1088} (\bibinfo{year}{2017}).
\newblock \url{https://www.nature.com/articles/nmat5017}.

\bibitem{Murray-bismuth2005}
\bibinfo{author}{Murray, E.~D.}, \bibinfo{author}{Fritz, D.~M.}, \bibinfo{author}{Wahlstrand, J.~K.}, \bibinfo{author}{Fahy, S.} \& \bibinfo{author}{Reis, D.~A.}
\newblock \bibinfo{title}{Effect of lattice anharmonicity on high-amplitude phonon dynamics in photoexcited bismuth}.
\newblock \emph{\bibinfo{journal}{Phys. Rev. B}} \textbf{\bibinfo{volume}{72}}, \bibinfo{pages}{060301} (\bibinfo{year}{2005}).
\newblock \url{https://link.aps.org/doi/10.1103/PhysRevB.72.060301}.

\bibitem{Forst-control2011}
\bibinfo{author}{F{\"o}rst, M.} \emph{et~al.}
\newblock \bibinfo{title}{Nonlinear phononics as an ultrafast route to lattice control}.
\newblock \emph{\bibinfo{journal}{Nat. Phys.}} \textbf{\bibinfo{volume}{7}}, \bibinfo{pages}{854--856} (\bibinfo{year}{2011}).
\newblock \url{https://doi.org/10.1038/nphys2055}.

\bibitem{ruello-review2015}
\bibinfo{author}{Ruello, P.} \& \bibinfo{author}{Gusev, V.~E.}
\newblock \bibinfo{title}{Physical mechanisms of coherent acoustic phonons generation by ultrafast laser action}.
\newblock \emph{\bibinfo{journal}{Ultrasonics}} \textbf{\bibinfo{volume}{56}}, \bibinfo{pages}{21--35} (\bibinfo{year}{2015}).
\newblock \url{https://www.sciencedirect.com/science/article/pii/S0041624X1400153X}.

\bibitem{Fausti2011light}
\bibinfo{author}{Fausti, D.} \emph{et~al.}
\newblock \bibinfo{title}{Light-induced superconductivity in a stripe-ordered cuprate}.
\newblock \emph{\bibinfo{journal}{Science}} \textbf{\bibinfo{volume}{331}}, \bibinfo{pages}{189--191} (\bibinfo{year}{2011}).
\newblock \url{https://www.science.org/doi/full/10.1126/science.1197294}.

\bibitem{Mitrano-Sc2016}
\bibinfo{author}{Mitrano, M.} \emph{et~al.}
\newblock \bibinfo{title}{Possible light-induced superconductivity in {K$_3$C$_{60}$} at high temperature}.
\newblock \emph{\bibinfo{journal}{Nature}} \textbf{\bibinfo{volume}{530}}, \bibinfo{pages}{461--464} (\bibinfo{year}{2016}).
\newblock \url{https://doi.org/10.1038/nature16522}.

\bibitem{Afanasiev-control2021}
\bibinfo{author}{Afanasiev, D.} \emph{et~al.}
\newblock \bibinfo{title}{Ultrafast control of magnetic interactions via light-driven phonons}.
\newblock \emph{\bibinfo{journal}{Nat. Mater.}} \textbf{\bibinfo{volume}{20}}, \bibinfo{pages}{607--611} (\bibinfo{year}{2021}).
\newblock \url{https://doi.org/10.1038/s41563-021-00922-7}.

\bibitem{Cavalleri-phase2004}
\bibinfo{author}{Cavalleri, A.}, \bibinfo{author}{Dekorsy, T.}, \bibinfo{author}{Chong, H. H.~W.}, \bibinfo{author}{Kieffer, J.~C.} \& \bibinfo{author}{Schoenlein, R.~W.}
\newblock \bibinfo{title}{Evidence for a structurally-driven insulator-to-metal transition in {VO$_2$}: A view from the ultrafast timescale}.
\newblock \emph{\bibinfo{journal}{Phys. Rev. B}} \textbf{\bibinfo{volume}{70}}, \bibinfo{pages}{161102} (\bibinfo{year}{2004}).
\newblock \url{https://link.aps.org/doi/10.1103/PhysRevB.70.161102}.

\bibitem{Kim-sdw2012}
\bibinfo{author}{Kim, K.~W.} \emph{et~al.}
\newblock \bibinfo{title}{Ultrafast transient generation of spin-density-wave order in the normal state of {BaFe$_2$As$_2$} driven by coherent lattice vibrations}.
\newblock \emph{\bibinfo{journal}{Nat. Mater.}} \textbf{\bibinfo{volume}{11}}, \bibinfo{pages}{497--501} (\bibinfo{year}{2012}).
\newblock \url{https://doi.org/10.1038/nmat3294}.

\bibitem{Hellmann-cdw2012}
\bibinfo{author}{Hellmann, S.} \emph{et~al.}
\newblock \bibinfo{title}{Time-domain classification of charge-density-wave insulators}.
\newblock \emph{\bibinfo{journal}{Nat. Commun.}} \textbf{\bibinfo{volume}{3}}, \bibinfo{pages}{1069} (\bibinfo{year}{2012}).
\newblock \url{https://doi.org/10.1038/ncomms2078}.

\bibitem{Katz-2012xc}
\bibinfo{author}{Katz, J.~E.} \emph{et~al.}
\newblock \bibinfo{title}{Electron small polarons and their mobility in iron (oxyhydr)oxide nanoparticles}.
\newblock \emph{\bibinfo{journal}{Science}} \textbf{\bibinfo{volume}{337}}, \bibinfo{pages}{1200--1203} (\bibinfo{year}{2012}).
\newblock \url{https://doi.org/10.1038/ncomms9420}.

\bibitem{Zurch-germanium2017}
\bibinfo{author}{Z{\"u}rch, M.} \emph{et~al.}
\newblock \bibinfo{title}{Direct and simultaneous observation of ultrafast electron and hole dynamics in germanium}.
\newblock \emph{\bibinfo{journal}{Nat. Commun.}} \textbf{\bibinfo{volume}{8}}, \bibinfo{pages}{15734} (\bibinfo{year}{2017}).
\newblock \url{https://doi.org/10.1038/ncomms15734}.

\bibitem{Cushing-hc2018}
\bibinfo{author}{Cushing, S.~K.} \emph{et~al.}
\newblock \bibinfo{title}{Hot phonon and carrier relaxation in {Si}(100) determined by transient extreme ultraviolet spectroscopy}.
\newblock \emph{\bibinfo{journal}{Struct. Dyn.}} \textbf{\bibinfo{volume}{5}}, \bibinfo{pages}{054302} (\bibinfo{year}{2018}).
\newblock \url{https://doi.org/10.1063/1.5038015}.

\bibitem{Mohr-inelaticxray2007}
\bibinfo{author}{Mohr, M.} \emph{et~al.}
\newblock \bibinfo{title}{Phonon dispersion of graphite by inelastic x-ray scattering}.
\newblock \emph{\bibinfo{journal}{Phys. Rev. B}} \textbf{\bibinfo{volume}{76}}, \bibinfo{pages}{035439} (\bibinfo{year}{2007}).
\newblock \url{https://link.aps.org/doi/10.1103/PhysRevB.76.035439}.

\bibitem{Trigo-xray2013}
\bibinfo{author}{Trigo, M.} \emph{et~al.}
\newblock \bibinfo{title}{Fourier-transform inelastic {X}-ray scattering from time- and momentum-dependent phonon--phonon correlations}.
\newblock \emph{\bibinfo{journal}{Nat. Phys.}} \textbf{\bibinfo{volume}{9}}, \bibinfo{pages}{790--794} (\bibinfo{year}{2013}).
\newblock \url{https://doi.org/10.1038/nphys2788}.

\bibitem{Valla-arpes1999}
\bibinfo{author}{Valla, T.}, \bibinfo{author}{Fedorov, A.~V.}, \bibinfo{author}{Johnson, P.~D.} \& \bibinfo{author}{Hulbert, S.~L.}
\newblock \bibinfo{title}{Many-body effects in angle-resolved photoemission: Quasiparticle energy and lifetime of a {Mo}(110) surface state}.
\newblock \emph{\bibinfo{journal}{Phys. Rev. Lett.}} \textbf{\bibinfo{volume}{83}}, \bibinfo{pages}{2085--2088} (\bibinfo{year}{1999}).
\newblock \url{https://link.aps.org/doi/10.1103/PhysRevLett.83.2085}.

\bibitem{Na-graphene2019}
\bibinfo{author}{Na, M.~X.} \emph{et~al.}
\newblock \bibinfo{title}{Direct determination of mode-projected electron-phonon coupling in the time domain}.
\newblock \emph{\bibinfo{journal}{Science}} \textbf{\bibinfo{volume}{366}}, \bibinfo{pages}{1231--1236} (\bibinfo{year}{2019}).
\newblock \url{https://www.science.org/doi/abs/10.1126/science.aaw1662}.
\newblock \eprint{https://www.science.org/doi/pdf/10.1126/science.aaw1662}.

\bibitem{Maznev-thermal2011}
\bibinfo{author}{Maznev, A.~A.}, \bibinfo{author}{Johnson, J.~A.} \& \bibinfo{author}{Nelson, K.~A.}
\newblock \bibinfo{title}{Onset of nondiffusive phonon transport in transient thermal grating decay}.
\newblock \emph{\bibinfo{journal}{Phys. Rev. B}} \textbf{\bibinfo{volume}{84}}, \bibinfo{pages}{195206} (\bibinfo{year}{2011}).
\newblock \url{https://link.aps.org/doi/10.1103/PhysRevB.84.195206}.

\bibitem{Baroni-DFT2001}
\bibinfo{author}{Baroni, S.}, \bibinfo{author}{de~Gironcoli, S.}, \bibinfo{author}{Dal~Corso, A.} \& \bibinfo{author}{Giannozzi, P.}
\newblock \bibinfo{title}{Phonons and related crystal properties from density-functional perturbation theory}.
\newblock \emph{\bibinfo{journal}{Rev. Mod. Phys.}} \textbf{\bibinfo{volume}{73}}, \bibinfo{pages}{515--562} (\bibinfo{year}{2001}).
\newblock \url{https://link.aps.org/doi/10.1103/RevModPhys.73.515}.

\bibitem{bernardi-theory2016}
\bibinfo{author}{Bernardi, M.}
\newblock \bibinfo{title}{First-principles dynamics of electrons and phonons}.
\newblock \emph{\bibinfo{journal}{Eur. Phys. J. B}} \textbf{\bibinfo{volume}{89}}, \bibinfo{pages}{1--15} (\bibinfo{year}{2016}).
\newblock \url{https://doi.org/10.1140/epjb/e2016-70399-4}.

\bibitem{Zhou-perturbo2021}
\bibinfo{author}{Zhou, J.-J.} \emph{et~al.}
\newblock \bibinfo{title}{Perturbo: A software package for ab initio electron--phonon interactions, charge transport and ultrafast dynamics}.
\newblock \emph{\bibinfo{journal}{Comput. Phys. Commun.}} \textbf{\bibinfo{volume}{264}}, \bibinfo{pages}{107970} (\bibinfo{year}{2021}).
\newblock \url{https://doi.org/10.1016/j.cpc.2021.107970}.

\bibitem{Giustino-Review2017}
\bibinfo{author}{Giustino, F.}
\newblock \bibinfo{title}{Electron-phonon interactions from first principles}.
\newblock \emph{\bibinfo{journal}{Rev. Mod. Phys.}} \textbf{\bibinfo{volume}{89}}, \bibinfo{pages}{015003} (\bibinfo{year}{2017}).
\newblock \url{https://link.aps.org/doi/10.1103/RevModPhys.89.015003}.

\bibitem{Bernardi-thermalization2014}
\bibinfo{author}{Bernardi, M.}, \bibinfo{author}{Vigil-Fowler, D.}, \bibinfo{author}{Lischner, J.}, \bibinfo{author}{Neaton, J.~B.} \& \bibinfo{author}{Louie, S.~G.}
\newblock \bibinfo{title}{Ab initio study of hot carriers in the first picosecond after sunlight absorption in silicon}.
\newblock \emph{\bibinfo{journal}{Phys. Rev. Lett.}} \textbf{\bibinfo{volume}{112}}, \bibinfo{pages}{257402} (\bibinfo{year}{2014}).
\newblock \url{https://link.aps.org/doi/10.1103/PhysRevLett.112.257402}.

\bibitem{Sangalli-photo2015}
\bibinfo{author}{Sangalli, D.} \& \bibinfo{author}{Marini, A.}
\newblock \bibinfo{title}{Ultra-fast carriers relaxation in bulk silicon following photo-excitation with a short and polarized laser pulse}.
\newblock \emph{\bibinfo{journal}{Europhy. Lett.}} \textbf{\bibinfo{volume}{110}}, \bibinfo{pages}{47004} (\bibinfo{year}{2015}).
\newblock \url{https://iopscience.iop.org/article/10.1209/0295-5075/110/47004/meta}.

\bibitem{Jhalani-BTE2017}
\bibinfo{author}{Jhalani, V.~A.}, \bibinfo{author}{Zhou, J.-J.} \& \bibinfo{author}{Bernardi, M.}
\newblock \bibinfo{title}{Ultrafast hot carrier dynamics in {G}a{N} and its impact on the efficiency droop}.
\newblock \emph{\bibinfo{journal}{Nano Lett.}} \textbf{\bibinfo{volume}{17}}, \bibinfo{pages}{5012--5019} (\bibinfo{year}{2017}).
\newblock \url{http://dx.doi.org/10.1021/acs.nanolett.7b02212}.

\bibitem{sjakste2018hot}
\bibinfo{author}{Sjakste, J.}, \bibinfo{author}{Tanimura, K.}, \bibinfo{author}{Barbarino, G.}, \bibinfo{author}{Perfetti, L.} \& \bibinfo{author}{Vast, N.}
\newblock \bibinfo{title}{Hot electron relaxation dynamics in semiconductors: assessing the strength of the electron--phonon coupling from the theoretical and experimental viewpoints}.
\newblock \emph{\bibinfo{journal}{J. Phys.: Condens. Matter}} \textbf{\bibinfo{volume}{30}}, \bibinfo{pages}{353001} (\bibinfo{year}{2018}).
\newblock \url{https://iopscience.iop.org/article/10.1088/1361-648X/aad487/}.

\bibitem{Zheng-theory2023}
\bibinfo{author}{Zheng, Z.} \emph{et~al.}
\newblock \bibinfo{title}{Ab initio real-time quantum dynamics of charge carriers in momentum space}.
\newblock \emph{\bibinfo{journal}{Nat. Comput. Sci.}} \textbf{\bibinfo{volume}{3}}, \bibinfo{pages}{532--541} (\bibinfo{year}{2023}).
\newblock \url{https://doi.org/10.1038\%2Fs43588-023-00456-9}.

\bibitem{Maliyov-BTE2021}
\bibinfo{author}{Maliyov, I.}, \bibinfo{author}{Park, J.} \& \bibinfo{author}{Bernardi, M.}
\newblock \bibinfo{title}{Ab initio electron dynamics in high electric fields: Accurate prediction of velocity-field curves}.
\newblock \emph{\bibinfo{journal}{Phys. Rev. B}} \textbf{\bibinfo{volume}{104}}, \bibinfo{pages}{L100303} (\bibinfo{year}{2021}).
\newblock \url{http://dx.doi.org/10.1103/PhysRevB.104.L100303}.

\bibitem{maliyov2024dynamic}
\bibinfo{author}{Maliyov, I.}, \bibinfo{author}{Yin, J.}, \bibinfo{author}{Yao, J.}, \bibinfo{author}{Yang, C.} \& \bibinfo{author}{Bernardi, M.}
\newblock \bibinfo{title}{Dynamic mode decomposition of nonequilibrium electron-phonon dynamics: accelerating the first-principles real-time {Boltzmann} equation}.
\newblock \emph{\bibinfo{journal}{npj Comput. Mater.}} \textbf{\bibinfo{volume}{10}}, \bibinfo{pages}{123} (\bibinfo{year}{2024}).
\newblock \url{https://www.nature.com/articles/s41524-024-01308-4}.

\bibitem{Tong-BTE2021}
\bibinfo{author}{Tong, X.} \& \bibinfo{author}{Bernardi, M.}
\newblock \bibinfo{title}{Toward precise simulations of the coupled ultrafast dynamics of electrons and atomic vibrations in materials}.
\newblock \emph{\bibinfo{journal}{Phys. Rev. Res.}} \textbf{\bibinfo{volume}{3}}, \bibinfo{pages}{023072} (\bibinfo{year}{2021}).
\newblock \url{https://link.aps.org/doi/10.1103/PhysRevResearch.3.023072}.

\bibitem{Caruso-BTE2021}
\bibinfo{author}{Caruso, F.}
\newblock \bibinfo{title}{Nonequilibrium lattice dynamics in monolayer {MoS$_2$}}.
\newblock \emph{\bibinfo{journal}{J. Phys. Chem. Lett.}} \textbf{\bibinfo{volume}{12}}, \bibinfo{pages}{1734--1740} (\bibinfo{year}{2021}).
\newblock \url{https://doi.org/10.1021\%2Facs.jpclett.0c03616}.

\bibitem{Caruso-chiral}
\bibinfo{author}{Pan, Y.} \& \bibinfo{author}{Caruso, F.}
\newblock \bibinfo{title}{Vibrational dichroism of chiral valley phonons}.
\newblock \emph{\bibinfo{journal}{Nano Lett.}} \textbf{\bibinfo{volume}{23}}, \bibinfo{pages}{7463--7469} (\bibinfo{year}{2023}).
\newblock \url{https://pubs.acs.org/doi/full/10.1021/acs.nanolett.3c01904}.

\bibitem{Prezhdo-NAMD2021}
\bibinfo{author}{Akimov, A.~V.}, \bibinfo{author}{Neukirch, A.~J.} \& \bibinfo{author}{Prezhdo, O.~V.}
\newblock \bibinfo{title}{Theoretical insights into photoinduced charge transfer and catalysis at oxide interfaces}.
\newblock \emph{\bibinfo{journal}{Chem. Rev.}} \textbf{\bibinfo{volume}{113}}, \bibinfo{pages}{4496--4565} (\bibinfo{year}{2013}).
\newblock \url{https://doi.org/10.1021/cr3004899}.

\bibitem{Perfetto-GW2022}
\bibinfo{author}{Perfetto, E.}, \bibinfo{author}{Pavlyukh, Y.} \& \bibinfo{author}{Stefanucci, G.}
\newblock \bibinfo{title}{Real-time {GW}: Toward an ab initio description of the ultrafast carrier and exciton dynamics in two-dimensional materials}.
\newblock \emph{\bibinfo{journal}{Phys. Rev. Lett.}} \textbf{\bibinfo{volume}{128}}, \bibinfo{pages}{016801} (\bibinfo{year}{2022}).
\newblock \url{https://doi.org/10.1103/PhysRevLett.128.016801}.

\bibitem{Perfetto2023real}
\bibinfo{author}{Perfetto, E.} \& \bibinfo{author}{Stefanucci, G.}
\newblock \bibinfo{title}{Real-time gw-ehrenfest-fan-migdal method for nonequilibrium 2{D} materials}.
\newblock \emph{\bibinfo{journal}{Nano Lett.}} \textbf{\bibinfo{volume}{23}}, \bibinfo{pages}{7029--7036} (\bibinfo{year}{2023}).
\newblock \url{https://pubs.acs.org/doi/full/10.1021/acs.nanolett.3c01772}.

\bibitem{Rozzi}
\bibinfo{author}{Andrea~Rozzi, C.} \emph{et~al.}
\newblock \bibinfo{title}{Quantum coherence controls the charge separation in a prototypical artificial light-harvesting system}.
\newblock \emph{\bibinfo{journal}{Nat. Commun.}} \textbf{\bibinfo{volume}{4}}, \bibinfo{pages}{1602} (\bibinfo{year}{2013}).
\newblock \url{https://www.nature.com/articles/ncomms2603}.

\bibitem{Maradudin-anharm1962}
\bibinfo{author}{Maradudin, A.~A.} \& \bibinfo{author}{Fein, A.~E.}
\newblock \bibinfo{title}{Scattering of neutrons by an anharmonic crystal}.
\newblock \emph{\bibinfo{journal}{Phys. Rev.}} \textbf{\bibinfo{volume}{128}}, \bibinfo{pages}{2589--2608} (\bibinfo{year}{1962}).
\newblock \url{https://link.aps.org/doi/10.1103/PhysRev.128.2589}.

\bibitem{Cowley-anharm1968}
\bibinfo{author}{Cowley, R.~A.}
\newblock \bibinfo{title}{Anharmonic crystals}.
\newblock \emph{\bibinfo{journal}{Rep. Prog. Phys.}} \textbf{\bibinfo{volume}{31}}, \bibinfo{pages}{123} (\bibinfo{year}{1968}).
\newblock \url{https://dx.doi.org/10.1088/0034-4885/31/1/303}.

\bibitem{Debernardi-anharm1995}
\bibinfo{author}{Debernardi, A.}, \bibinfo{author}{Baroni, S.} \& \bibinfo{author}{Molinari, E.}
\newblock \bibinfo{title}{Anharmonic phonon lifetimes in semiconductors from density-functional perturbation theory}.
\newblock \emph{\bibinfo{journal}{Phys. Rev. Lett.}} \textbf{\bibinfo{volume}{75}}, \bibinfo{pages}{1819--1822} (\bibinfo{year}{1995}).
\newblock \url{https://link.aps.org/doi/10.1103/PhysRevLett.75.1819}.

\bibitem{Li-ShengBTE2014}
\bibinfo{author}{Li, W.}, \bibinfo{author}{Carrete, J.}, \bibinfo{author}{Katcho, N.~A.} \& \bibinfo{author}{Mingo, N.}
\newblock \bibinfo{title}{{ShengBTE:} a solver of the {B}oltzmann transport equation for phonons}.
\newblock \emph{\bibinfo{journal}{Comp. Phys. Commun.}} \textbf{\bibinfo{volume}{185}}, \bibinfo{pages}{1747–1758} (\bibinfo{year}{2014}).
\newblock \url{https://doi.org/10.1016/j.cpc.2014.02.015}.

\bibitem{Atsushi-phono3py2015}
\bibinfo{author}{Togo, A.}, \bibinfo{author}{Chaput, L.} \& \bibinfo{author}{Tanaka, I.}
\newblock \bibinfo{title}{Distributions of phonon lifetimes in {Brillouin} zones}.
\newblock \emph{\bibinfo{journal}{Phys. Rev. B}} \textbf{\bibinfo{volume}{91}}, \bibinfo{pages}{094306} (\bibinfo{year}{2015}).
\newblock \url{https://link.aps.org/doi/10.1103/PhysRevB.91.094306}.

\bibitem{carrete-almaBTE2017}
\bibinfo{author}{Carrete, J.} \emph{et~al.}
\newblock \bibinfo{title}{{almaBTE: A} solver of the space–time dependent {Boltzmann} transport equation for phonons in structured materials}.
\newblock \emph{\bibinfo{journal}{Comput. Phys. Commun.}} \textbf{\bibinfo{volume}{220}}, \bibinfo{pages}{351--362} (\bibinfo{year}{2017}).
\newblock \url{https://www.sciencedirect.com/science/article/pii/S0010465517302059}.

\bibitem{Britt-BTE2022}
\bibinfo{author}{Britt, T.~L.} \emph{et~al.}
\newblock \bibinfo{title}{Direct view of phonon dynamics in atomically thin {MoS$_2$}}.
\newblock \emph{\bibinfo{journal}{Nano Lett.}} \textbf{\bibinfo{volume}{22}}, \bibinfo{pages}{4718--4724} (\bibinfo{year}{2022}).
\newblock \url{https://doi.org/10.1021/acs.nanolett.2c00850}.

\bibitem{hindmarsh2005sundials}
\bibinfo{author}{Hindmarsh, A.~C.} \emph{et~al.}
\newblock \bibinfo{title}{{SUNDIALS}: Suite of nonlinear and differential/algebraic equation solvers}.
\newblock \emph{\bibinfo{journal}{ACM Trans. Math. Softw.}} \textbf{\bibinfo{volume}{31}}, \bibinfo{pages}{363--396} (\bibinfo{year}{2005}).
\newblock \url{https://doi.org/10.1145/1089014.1089020}.

\bibitem{gardner2022enabling}
\bibinfo{author}{Gardner, D.~J.}, \bibinfo{author}{Reynolds, D.~R.}, \bibinfo{author}{Woodward, C.~S.} \& \bibinfo{author}{Balos, C.~J.}
\newblock \bibinfo{title}{Enabling new flexibility in the {SUNDIALS} suite of nonlinear and differential/algebraic equation solvers}.
\newblock \emph{\bibinfo{journal}{ACM Trans. Math. Softw.}} \textbf{\bibinfo{volume}{48}}, \bibinfo{pages}{1--24} (\bibinfo{year}{2022}).
\newblock \url{https://doi.org/10.1145/3539801}.

\bibitem{reynolds2023arkode}
\bibinfo{author}{Reynolds, D.~R.}, \bibinfo{author}{Gardner, D.~J.}, \bibinfo{author}{Woodward, C.~S.} \& \bibinfo{author}{Chinomona, R.}
\newblock \bibinfo{title}{{ARKODE}: A flexible {IVP} solver infrastructure for one-step methods}.
\newblock \emph{\bibinfo{journal}{ACM Trans. Math. Softw.}} \textbf{\bibinfo{volume}{49}}, \bibinfo{pages}{1--26} (\bibinfo{year}{2023}).
\newblock \url{https://doi.org/10.1145/3594632}.

\bibitem{hairer2008solving}
\bibinfo{author}{Harier, E.}, \bibinfo{author}{N{\o}rsett, S.~P.} \& \bibinfo{author}{Wanner, G.}
\newblock \emph{\bibinfo{title}{Solving Ordinary Differential Equations I: Nonstiff Problems}}.
\newblock Springer Series in Computational Mathematics (\bibinfo{publisher}{Springer}, \bibinfo{year}{2008}), \bibinfo{edition}{2} edn.
\newblock \url{https://doi.org/10.1007/978-3-540-78862-1}.

\bibitem{sandu2019class}
\bibinfo{author}{Sandu, A.}
\newblock \bibinfo{title}{A class of multirate infinitesimal {GARK} methods}.
\newblock \emph{\bibinfo{journal}{SIAM J. Numer. Anal.}} \textbf{\bibinfo{volume}{57}}, \bibinfo{pages}{2300--2327} (\bibinfo{year}{2019}).
\newblock \url{https://doi.org/10.1137/18M1205492}.

\bibitem{schlegel2009multirate}
\bibinfo{author}{Schlegel, M.}, \bibinfo{author}{Knoth, O.}, \bibinfo{author}{Arnold, M.} \& \bibinfo{author}{Wolke, R.}
\newblock \bibinfo{title}{Multirate {Runge--Kutta} schemes for advection equations}.
\newblock \emph{\bibinfo{journal}{J. Comput. Appl. Math.}} \textbf{\bibinfo{volume}{226}}, \bibinfo{pages}{345--357} (\bibinfo{year}{2009}).
\newblock \url{https://doi.org/10.1016/j.cam.2008.08.009}.

\bibitem{loffeld2024performance}
\bibinfo{author}{Loffeld, J.~J.}, \bibinfo{author}{Nonaka, A.}, \bibinfo{author}{Reynolds, D.~R.}, \bibinfo{author}{Gardner, D.~J.} \& \bibinfo{author}{Woodward, C.~S.}
\newblock \bibinfo{title}{Performance of explicit and {IMEX MRI} multirate methods on complex reactive flow problems within modern parallel adaptive structured grid frameworks}.
\newblock \emph{\bibinfo{journal}{Int. J. High Perform. Comput. Appl.}} \textbf{\bibinfo{volume}{38}}, \bibinfo{pages}{263--281} (\bibinfo{year}{2024}).
\newblock \url{https://doi.org/10.1177/10943420241227914}.

\bibitem{Holt-tds1999}
\bibinfo{author}{Holt, M.} \emph{et~al.}
\newblock \bibinfo{title}{Determination of phonon dispersions from {X}-ray transmission scattering: The example of silicon}.
\newblock \emph{\bibinfo{journal}{Phys. Rev. Lett.}} \textbf{\bibinfo{volume}{83}}, \bibinfo{pages}{3317--3319} (\bibinfo{year}{1999}).
\newblock \url{https://link.aps.org/doi/10.1103/PhysRevLett.83.3317}.

\bibitem{Filippetto-TDSreview2022}
\bibinfo{author}{Filippetto, D.} \emph{et~al.}
\newblock \bibinfo{title}{Ultrafast electron diffraction: Visualizing dynamic states of matter}.
\newblock \emph{\bibinfo{journal}{Rev. Mod. Phys.}} \textbf{\bibinfo{volume}{94}}, \bibinfo{pages}{045004} (\bibinfo{year}{2022}).
\newblock \url{https://link.aps.org/doi/10.1103/RevModPhys.94.045004}.

\bibitem{Warren-xray1990}
\bibinfo{author}{Warren, B.~E.}
\newblock \emph{\bibinfo{title}{{X}-Ray Diffraction}} (\bibinfo{publisher}{Dover Publications}, \bibinfo{address}{New York}, \bibinfo{year}{1990}).

\bibitem{Chen-exciton2020}
\bibinfo{author}{Chen, H.-Y.}, \bibinfo{author}{Sangalli, D.} \& \bibinfo{author}{Bernardi, M.}
\newblock \bibinfo{title}{Exciton-phonon interaction and relaxation times from first principles}.
\newblock \emph{\bibinfo{journal}{Phys. Rev. Lett.}} \textbf{\bibinfo{volume}{125}}, \bibinfo{pages}{107401} (\bibinfo{year}{2020}).
\newblock \url{https://link.aps.org/doi/10.1103/PhysRevLett.125.107401}.

\bibitem{Chen-exciton2022}
\bibinfo{author}{Chen, H.-Y.}, \bibinfo{author}{Sangalli, D.} \& \bibinfo{author}{Bernardi, M.}
\newblock \bibinfo{title}{First-principles ultrafast exciton dynamics and time-domain spectroscopies: Dark-exciton mediated valley depolarization in monolayer ${\mathrm{wse}}_{2}$}.
\newblock \emph{\bibinfo{journal}{Phys. Rev. Res.}} \textbf{\bibinfo{volume}{4}}, \bibinfo{pages}{043203} (\bibinfo{year}{2022}).
\newblock \url{https://link.aps.org/doi/10.1103/PhysRevResearch.4.043203}.

\bibitem{Perfetto-exciton2024}
\bibinfo{author}{Perfetto, E.}, \bibinfo{author}{Wu, K.} \& \bibinfo{author}{Stefanucci, G.}
\newblock \bibinfo{title}{Theory of coherent phonons coupled to excitons}.
\newblock \emph{\bibinfo{journal}{npj 2D Mater. Appl.}} \textbf{\bibinfo{volume}{8}}, \bibinfo{pages}{40} (\bibinfo{year}{2024}).
\newblock \url{https://doi.org/10.1038/s41699-024-00474-9}.

\bibitem{Rosati-lindblad2014}
\bibinfo{author}{Rosati, R.}, \bibinfo{author}{Iotti, R.~C.}, \bibinfo{author}{Dolcini, F.} \& \bibinfo{author}{Rossi, F.}
\newblock \bibinfo{title}{Derivation of nonlinear single-particle equations via many-body lindblad superoperators: A density-matrix approach}.
\newblock \emph{\bibinfo{journal}{Phys. Rev. B}} \textbf{\bibinfo{volume}{90}}, \bibinfo{pages}{125140} (\bibinfo{year}{2014}).
\newblock \url{https://link.aps.org/doi/10.1103/PhysRevB.90.125140}.

\bibitem{Martin-DFT2020}
\bibinfo{author}{Martin, R.~M.}
\newblock \emph{\bibinfo{title}{Electronic Structure: Basic Theory and Practical Methods}} (\bibinfo{publisher}{Cambridge University Press}, \bibinfo{year}{2020}).
\newblock \url{https://doi.org/10.1017/CBO9780511805769}.

\bibitem{QE2017}
\bibinfo{author}{Giannozzi, P.} \emph{et~al.}
\newblock \bibinfo{title}{Advanced capabilities for materials modelling with {QUANTUM ESPRESSO}}.
\newblock \emph{\bibinfo{journal}{J. Phys. Condens. Matter}} \textbf{\bibinfo{volume}{29}}, \bibinfo{pages}{465901} (\bibinfo{year}{2017}).
\newblock \url{https://doi.org/10.1088/1361-648X/aa8f79}.

\bibitem{Mostofi-wannier2014}
\bibinfo{author}{Mostofi, A.~A.} \emph{et~al.}
\newblock \bibinfo{title}{An updated version of {WANNIER90}: A tool for obtaining maximally-localised {Wannier} functions}.
\newblock \emph{\bibinfo{journal}{Comput. Phys. Commun.}} \textbf{\bibinfo{volume}{185}}, \bibinfo{pages}{2309 -- 2310} (\bibinfo{year}{2014}).
\newblock \url{https://doi.org/10.1016/j.cpc.2014.05.003}.

\bibitem{mahan-condensed2010}
\bibinfo{author}{Mahan, G.~D.}
\newblock \emph{\bibinfo{title}{Condensed Matter in a Nutshell}} (\bibinfo{publisher}{Princeton University Press}, \bibinfo{year}{2010}).

\bibitem{Hellman-tdep2013}
\bibinfo{author}{Hellman, O.} \& \bibinfo{author}{Abrikosov, I.~A.}
\newblock \bibinfo{title}{Temperature-dependent effective third-order interatomic force constants from first principles}.
\newblock \emph{\bibinfo{journal}{Phys. Rev. B}} \textbf{\bibinfo{volume}{88}}, \bibinfo{pages}{144301} (\bibinfo{year}{2013}).
\newblock \url{https://link.aps.org/doi/10.1103/PhysRevB.88.144301}.

\bibitem{Zhou-STO2018}
\bibinfo{author}{Zhou, J.-J.}, \bibinfo{author}{Hellman, O.} \& \bibinfo{author}{Bernardi, M.}
\newblock \bibinfo{title}{Electron-phonon scattering in the presence of soft modes and electron mobility in {SrTiO$_{3}$} perovskite from first principles}.
\newblock \emph{\bibinfo{journal}{Phys. Rev. Lett.}} \textbf{\bibinfo{volume}{121}}, \bibinfo{pages}{226603} (\bibinfo{year}{2018}).
\newblock \url{https://link.aps.org/doi/10.1103/PhysRevLett.121.226603}.

\bibitem{Olukayode-xray2023}
\bibinfo{author}{Olukayode, S.}, \bibinfo{author}{Froese~Fischer, C.} \& \bibinfo{author}{Volkov, A.}
\newblock \bibinfo{title}{{Revisited relativistic Dirac--Hartree--Fock {X}-ray scattering factors. I. Neutral atoms with Z = 2--118}}.
\newblock \emph{\bibinfo{journal}{Acta Crystallogr. Sect. A}} \textbf{\bibinfo{volume}{79}}, \bibinfo{pages}{59--79} (\bibinfo{year}{2023}).
\newblock \url{https://doi.org/10.1107/S2053273322010944}.

\bibitem{Zonneveld1963automatic}
\bibinfo{author}{Zonneveld, J.}
\newblock \bibinfo{title}{Automatic integration of ordinary differential equations}.
\newblock \bibinfo{type}{Tech. Rep.} \bibinfo{number}{R743}, \bibinfo{institution}{Mathematisch Centrum} (\bibinfo{year}{1963}).
\newblock \url{https://doi.org/10.1145/362566.362571}.

\bibitem{soderlind1998automatic}
\bibinfo{author}{S{\"o}derlind, G.}
\newblock \bibinfo{title}{The automatic control of numerical integration}.
\newblock \emph{\bibinfo{journal}{CWI Quarterly}} \textbf{\bibinfo{volume}{11}}, \bibinfo{pages}{55--74} (\bibinfo{year}{1998}).

\bibitem{soderlind2003digital}
\bibinfo{author}{S{\"o}derlind, G.}
\newblock \bibinfo{title}{Digital filters in adaptive time-stepping}.
\newblock \emph{\bibinfo{journal}{ACM Trans. Math. Softw.}} \textbf{\bibinfo{volume}{29}}, \bibinfo{pages}{1--26} (\bibinfo{year}{2003}).
\newblock \url{https://doi.org/10.1145/641876.641877}.

\bibitem{soderlind2006time}
\bibinfo{author}{S{\"o}derlind, G.}
\newblock \bibinfo{title}{Time-step selection algorithms: Adaptivity, control, and signal processing}.
\newblock \emph{\bibinfo{journal}{Appl. Numer. Math.}} \textbf{\bibinfo{volume}{56}}, \bibinfo{pages}{488--502} (\bibinfo{year}{2006}).
\newblock \url{https://doi.org/10.1016/j.apnum.2005.04.026}.

\bibitem{schlegel2012implementation}
\bibinfo{author}{Schlegel, M.}, \bibinfo{author}{Knoth, O.}, \bibinfo{author}{Arnold, M.} \& \bibinfo{author}{Wolke, R.}
\newblock \bibinfo{title}{Implementation of multirate time integration methods for air pollution modelling}.
\newblock \emph{\bibinfo{journal}{Geosci. Model Dev.}} \textbf{\bibinfo{volume}{5}}, \bibinfo{pages}{1395--1405} (\bibinfo{year}{2012}).
\newblock \url{https://doi.org/10.5194/gmd-5-1395-2012, 2012}.

\bibitem{schlegel2012numerical}
\bibinfo{author}{Schlegel, M.}, \bibinfo{author}{Knoth, O.}, \bibinfo{author}{Arnold, M.} \& \bibinfo{author}{Wolke, R.}
\newblock \bibinfo{title}{Numerical solution of multiscale problems in atmospheric modeling}.
\newblock \emph{\bibinfo{journal}{Appl. Numer. Math.}} \textbf{\bibinfo{volume}{62}}, \bibinfo{pages}{1531--1543} (\bibinfo{year}{2012}).
\newblock \url{https://doi.org/10.1016/j.apnum.2012.06.023}.

\bibitem{knoth1998implicit}
\bibinfo{author}{Knoth, O.} \& \bibinfo{author}{Wolke, R.}
\newblock \bibinfo{title}{Implicit-explicit {Runge-Kutta} methods for computing atmospheric reactive flows}.
\newblock \emph{\bibinfo{journal}{Appl. Numer. Math.}} \textbf{\bibinfo{volume}{28}}, \bibinfo{pages}{327--341} (\bibinfo{year}{1998}).
\newblock \url{https://doi.org/10.1016/S0168-9274(98)00051-8}.

\bibitem{Swig-Fortran}
\bibinfo{author}{Johnson, S.~R.}, \bibinfo{author}{Prokopenko, A.} \& \bibinfo{author}{Evans, K.~J.}
\newblock \bibinfo{title}{Automated fortran-\rm{C}++ bindings for large-scale scientific applications}.
\newblock \emph{\bibinfo{journal}{Comput. Sci. Eng.}} \textbf{\bibinfo{volume}{22}}, \bibinfo{pages}{84--94} (\bibinfo{year}{2020}).
\newblock \url{https://doi.org/10.1109/MCSE.2019.2924204}.

\end{thebibliography}
\end{document}